\documentclass
[aps,preprint,amsfonts,amssymb,showkeys,nobibnotes,nofootinbib,oneside,titlepage,11pt,groupedaddress,noshowpacs]{revtex4}%
\usepackage{amsfonts}
\usepackage{amsmath}
\usepackage{amssymb}
\usepackage{graphicx}
\usepackage{natbib}%
\providecommand{\U}[1]{\protect\rule{.1in}{.1in}}

\makeatletter
\def\Dated@name{}
\makeatother
\begin{document}
\title[Heisenberg groups over algebras]{Heisenberg groups and their automorphisms over algebras with central
involution }
\author{Robert W. Johnson}
\date{email: \ rwjcontact@aol.com}
\begin{abstract}
Heisenberg groups over algebras with central involution and their
\ automorphism groups are constructed. \ The complex quaternion group algebra
over a prime field is used as an example. Its subspaces provide finite models
for each of the real and complex quadratic spaces with dimension 4 or less.
\ \ A model for the representations of these Heisenberg groups and
automorphism groups is constructed. \ A pseudo-differential operator enables a
parallel treatment of spaces defined over finite and real fields.

\end{abstract}
\maketitle

\section{Introduction}

\bigskip Our goal in this paper is to construct particular finite groups along
with their representations that can be used as models for physical systems in
dimensions greater than 1. \ The interrelationship between quantum mechanics
and group theory is an old and vast subject. \ Recent developments in
constucting Weil representations for SL$_{\ast}(2,A)$ for quite general
involutive algebras $A$ \cite{Gu1} and new insights into the relationship
between the Jacobi group, that is formed by the semidirect product of
SL$_{\ast}(2,A)$ with appropriate Heisenberg group, and the wavefunction and
Wigner distribution of associated quantum systems \cite{Ib,J2} motivate this
paper. \ \ To simplify this treatment we work with algebras defined over
finite fields. \ The finite Heisenberg groups that are formed over these
algebras admit an elementary and constructive treatment.

\bigskip Our starting point is an algebra $A$ with central involution $\ast$
that is used to define a trace and norm. \ We use elements from this algebra
as entries in a 4x4 matrix presentation of a Heisenberg group H$_{\ast}(A).$
\ This presentation extends the treatment of Berndt and Schmidt (\cite{BS} p
2) of Heisenberg groups over a commutative ring. \ 

\ Groups of automorphisms of H$_{\ast}(A)$ are then introduced. \ We recover
the automorphism group SL$_{\ast}(2,A)$ described by Pantoja and Soto-Andrade
\cite{Pn}. \ For orientation,\ SL$_{\ast}(2,A)$ reduces to the symplectic
group Sp$(2n,F)$ when $A$ is the ring of matices M$_{n}(F)$ over a field $F$
and involution corresponds to matrix transpose. \ \ We then consider the
semidirect product SL$_{\ast}(2,A)\ltimes$H$_{\ast}(A)$ that is the Jacobi group.

The Schr\"{o}dinger representation\ for H$_{\ast}(A)$ is readily constructed
and is unique for fixed nonzero central character. \ The Weil representation
of SL$_{\ast}(2,A)$ is associated with automorphisms of the Schr\"{o}dinger
representation. \ Gutierrez, Pantoja and Soto-Andrade \cite{Gu1} show how Weil
representations for SL$_{\ast}(2,A)$ can be constructed when SL$_{\ast}(2,A)$
admits a Bruhat decomposition. \ Using this information one may then construct
the Schr\"{o}dinger-Weil representation for the semidirect product SL$_{\ast
}(2,A)\ltimes$H$_{\ast}(A)$ for a wide range of algebras. \ 

We may identify a function on the group algebra of H$_{\ast}(A)$ that
transforms according to the Schr\"{o}dinger representation with the
wavefunction of a quantum system whose configuration space is $A$. \ The
dynamical evolution of the wavefunction under SL$_{\ast}(2,A)$ automorphisms
is described by the Schr\"{o}dinger-Weil representation of SL$_{\ast
}(2,A)\ltimes$H$_{\ast}(A)$ (see, de Gosson \cite{deG},\ ch 15 ). \ From the
wavefunction one may then construct the Wigner distribution for the quantum
system \cite{Ib,J2,Kr}.

\bigskip Additional motivation for this paper includes the wish, addressed by
Low \cite{Lo1}, to treat relativistic quantum mechanics, with focus the
Poincare group, and non-relativistic quantum mechanics, with focus the
Heisenberg group, in a more uniform way. \ The work that we describe below
provides a framework, different from Low's, that combines these two cases in a
natural way and that points towards further generalization.

\bigskip Section II reviews some properties of the involutive algebras that
are used as matix entries in the construction of the Heisenberg groups
H$_{\ast}(A)$ in Sec. IIIA. \ In Sec. IIIB we construct the group SL$_{\ast
}(2,A)$ that preserves H$_{\ast}(A)$ under similarity transformation. \ When
we restrict from H$_{\ast}(A)$ to H$_{\ast}(S)$ for $S\subset A$ we must also
restrict SL$_{\ast}(2,A)$ to the subgroup that preserves H$_{\ast}(S)$ under
automorphisms. \ In this paper we consider in detail the case SL$(2,F_{N})$
where $F_{N}$ is the identity component of $A$. \ It is applicable to each of
the Heisenberg groups that we consider. \ In Sec. IIIC we construct the
diagonal subgroup D$(A_{0}^{x})$ whose entries\ are from the set of norm 1
elements of $A.$ \ Elements from\ D$(A_{0}^{x})$ generate norm preserving
transformations of the translational subgroups of H$_{\ast}(A).$ \ The group
relations are summarized in Sec. IIID. \ In Sec. IIIE we introduce the complex
quaternion group $C4\otimes Q$ and its algebra $C4\otimes Q(F_{N})$over the
finite field $F_{N}.$ \ Vector subspaces or subalgebras of $C4\otimes
Q(F_{N})$ provide finite models for each of the real or complex quadratic
spaces with dimension 4 or less.

In Sec. IV we construct representations for H$_{\ast}(A)$ and its automorphism
groups for $A=C4\otimes Q(F_{N}).$ \ In Sec. IVA, while defining notation, we
treat the representations of important cyclic subgroups. \ Representations of
the Heisenberg group H$_{\ast}(A)$ are described in Sec. IVB. \ In Sec. IVC we
consider a subspace of the group algebra of SL$(2,F_{N})\ltimes$H$_{\ast}(A)$
and show by direct calculation that it is closed under the action of the
group. \ Details of this calculation are contained in Appendix A. \ Vectors in
this subspace transform according to the Schr\"{o}dinger-Weil representation.
\ In Sec. IVD we describe how the Schr\"{o}dinger-Weil representation can be
extended to obtain representations of $($D$(A_{0}^{x})\otimes$SL$(2,F_{N}%
))\ltimes$H$_{\ast}(A).$

\bigskip We define a pseudo-differential operator\ in Sec. VA that allows one
to express group actions in a form that parallels the familar setting of
differential operators acting on functions defined over real spaces. \ In this
way, while avoiding complications having a purely analytical origin by working
over finite fields and rings, we may still gain insight into the structure of
the physical case. \ These pseudo-differential operators are also needed when
Wigner distributions for quantum systems based on these groups and
representations are constructed in future work. \ In Sec. VB we consider
eigenfunctions of particular group generators in the Schr\"{o}dinger-Weil
representation. \ 

\bigskip

\section{\bigskip Algebra with involution}

We consider Heisenberg groups whose group elements may be expressed as
matrices with entries from an algebra $A$ with involution. \ An involution is
a map $^{\ast}:A\rightarrow A$ subject to the following conditions
\begin{align*}
x^{\ast\ast}  &  =x\text{ }\\
(xy)^{\ast}  &  =y^{\ast}x^{\ast}\\
(x+y)^{\ast}  &  =x^{\ast}+y^{\ast}%
\end{align*}
for $x,y\in A$ (\cite{Kn} p 13 and \cite{McC} p 139). \ Elements that are
invariant under the involution are termed Hermitian and can be constructed in
2 standard ways \
\begin{align*}
tr(x)  &  =x+x^{\ast}\\
n(x)  &  =xx^{\ast}%
\end{align*}
where tr$(x)$ is the trace mapping and $n(x)$ is the norm map. \ For central
involutions the center of the algebra is Hermitian. \ For central involutions
the norm map can be composed%
\begin{align*}
n(xy)  &  =(xy)(xy)^{\ast}\\
&  =(xy)(y^{\ast}x^{\ast})\\
&  =n(x)n(y)
\end{align*}
When $n(x)$ is invertible then $x$ is also invertible%
\[
x^{-1}=x^{\ast}n(x)^{-1}%
\]

When the center of the algebra is multidimensional, we denote the identity
component of the trace as $Tr(x)$ and the identity compoenent of the norm as
$N(x).$ \ 

We restrict to associative algebras in this paper since these provide examples
that are already of physical interest and allow a simpler treatment. \ 

\section{Group Construction}

\subsection{\bigskip Heisenberg group}

\bigskip We consider the group that is generated by the matrices
\begin{equation}
t_{x}^{p}=\left(
\begin{array}
[c]{cccc}%
1 & 0 & 0 & 0\\
p & 1 & 0 & 0\\
0 & 0 & 1 & \varepsilon p^{\ast}\\
0 & 0 & 0 & 1
\end{array}
\right)  ;t_{y}^{q}=\left(
\begin{array}
[c]{cccc}%
1 & 0 & 0 & q^{\ast}\\
0 & 1 & q & 0\\
0 & 0 & 1 & 0\\
0 & 0 & 0 & 1
\end{array}
\right)  \label{h generators}%
\end{equation}
by matrix multiplication, where $p,q$ are elements in an algebra $A$ with
involution and $\varepsilon=\pm1.$ \ We call this the Heisenberg group
H$_{\ast}^{\varepsilon}(A).$ \ We have the relations%
\begin{align*}
t_{x}^{p}t_{x}^{p^{\prime}}  &  =t_{x}^{p+p^{\prime}}\\
t_{y}^{q}t_{y}^{q^{\prime}}  &  =t_{y}^{q+q^{\prime}}%
\end{align*}
We have the commutator%
\begin{equation}
t_{x}^{p}t_{y}^{q}t_{x}^{-p}t_{y}^{-q}=t_{z}^{pq^{\ast}-\varepsilon(pq^{\ast
})^{\ast}} \label{h commutator}%
\end{equation}
for
\[
t_{z}^{pq^{\ast}-\varepsilon(pq^{\ast})^{\ast}}=^{\text{def}}\left(
\begin{array}
[c]{cccc}%
1 & 0 & 0 & 0\\
0 & 1 & 0 & pq^{\ast}-\varepsilon(pq^{\ast})^{\ast}\\
0 & 0 & 1 & 0\\
0 & 0 & 0 & 1
\end{array}
\right)  \text{ }\
\]
The subgroup with elements $\{t_{z}^{t-\varepsilon t^{\ast}}$for $t\in A\}$
forms the center of H$_{\ast}^{\varepsilon}(A).$ \ When $\varepsilon=-1$ the
argument of $t_{z}$ is hermitian with respect to $\ast.$ \ This is the case of
most interest in this paper and in the following we restrict to $\varepsilon
=-1$. \ \ We donote this group H$_{\ast}(A).$ \ The group multiplication can
be written%
\begin{align*}
&  \left(  t_{x}^{p}t_{y}^{q}t_{z}^{\text{tr}(t)}\right)  \left(
t_{x}^{p^{\prime}}t_{y}^{q^{\prime}}t_{z}^{\text{tr}(t^{\prime})}\right) \\
&  =t_{x}^{p+p^{\prime}}t_{y}^{q+q^{\prime}}t_{z}^{\text{tr}(t+t^{\prime
}-qp^{\prime\ast})}%
\end{align*}

We may relax the requirement that $p,q\in$ algebra. \ The Heisenberg group
H$_{\ast}(S)$ is still well-defined if we require, less stringently, that
$p,q$ reside in a subspace $S\subset A$ that is closed under translations.
\ The argument of $t_{z}$ then resides in the subspace whose terms can be
written tr$(xy)$ for $x,y\in S$.

\bigskip Let us review the familiar construction of the Heisenberg group
H$(V)$ (see, e.g., Neuhauser\cite{N} and Folland \cite{F1} p 19) over the
vector space $V$ whose elements $v$ consist of pairs of vectors $(p,q)$
defined over a ground field $F$ and equipped with a symplectic form
$\left\langle p,q^{\prime}\right\rangle -\left\langle q,p^{\prime
}\right\rangle ,$ where $\left\langle p,q^{\prime}\right\rangle $ denotes a
nondegenerate symmetric bilinear form on the vector space that contains $p$
and $q^{\prime}.$ \ This Heisenberg group contains elements $(p,q,t)$ for
$t\in F$ equipped with the multiplication%
\[
(p,q,t)(p^{\prime},q^{\prime},t^{\prime})=(p+p^{\prime},q+q^{\prime
},t+t^{\prime}+\frac{1}{2}(\left\langle p,q^{\prime}\right\rangle
-\left\langle q,p^{\prime}\right\rangle ))\text{ }%
\]

Now consider the Heisenberg group H$_{\ast}(A)$ of this paper. \ Identify the
group element%
\[
t_{x}^{p}t_{y}^{q}t_{z}^{\text{tr}(t-\frac{1}{2}pq^{\ast})}\in H_{\ast}(A)
\]
with a triple $(p,q,$tr$(t))$ and obtain the multiplication%
\[
(p,q,\text{tr}(t))(p^{\prime},q^{\prime},\text{tr}(t^{\prime}))=(p+p^{\prime
},q+q^{\prime},\text{tr}(t+t^{\prime}+\frac{1}{2}(pq^{\prime\ast}%
-qp^{\prime\ast})))
\]
Our construction reduces to the standard construction when tr$(pq^{\ast})=$
$\left\langle p,q\right\rangle $.

\subsection{\bigskip Semidirect product with SL$_{\ast}(2,A)$}

\bigskip We now construct the group SL$_{\ast}(2,A)$ described by Pantoja and
Soto-Andrade \cite{Pn} as a group of automorphisms of H$_{\ast}(A)$. \ Pantoja
and Soto-Andrade \cite{Pn2} also consider the generalized case SL$_{\ast
}^{\varepsilon}(2,A)$ for $\varepsilon=\pm1.$ \ One can show that SL$_{\ast
}^{\varepsilon}(2,A)$ is an automorphism group of the generalized Heisenberg
group H$_{\ast}^{\varepsilon}(A).$ \ 

For $a,b,c,d\in A$ consider%
\[
g(a,b,c,d)=\left(
\begin{array}
[c]{cccc}%
a & 0 & b & 0\\
0 & 1 & 0 & 0\\
c & 0 & d & 0\\
0 & 0 & 0 & 1
\end{array}
\right)
\]
with inverse%
\[
g(a,b,c,d)^{-1}=g(x,y,z,w)
\]
for some $x,y,z,w\in A.$ \ We then have
\begin{align}
1  &  =ax+bz=cy+dw\nonumber\\
0  &  =ay+bw=cx+dz \label{invcond}%
\end{align}
Now consider the similarity transformation%
\begin{align*}
&  g(a,b,c,d)\left(
\begin{array}
[c]{cccc}%
1 & 0 & 0 & q^{\ast}\\
p & 1 & q & \text{tr}(t)+pq^{\ast}\\
0 & 0 & 1 & -p^{\ast}\\
0 & 0 & 0 & 1
\end{array}
\right)  g(a,b,c,d)^{-1}\\
&  =\left(
\begin{array}
[c]{cccc}%
1 & 0 & 0 & aq^{\ast}-bp^{\ast}\\
px+qz & 1 & py+qw & tr(t)+pq^{\ast}\\
0 & 0 & 1 & cq^{\ast}-dp^{\ast}\\
0 & 0 & 0 & 1
\end{array}
\right)
\end{align*}
Closure of H$_{\ast}(A)$ requires%
\begin{align*}
-\left(  px+qz\right)  ^{\ast}  &  =cq^{\ast}-dp^{\ast}\\
\left(  py+qw\right)  ^{\ast}  &  =aq^{\ast}-bp^{\ast}%
\end{align*}
for each $p,q.$ $\ $This$\ $implies%
\[
x^{\ast}=d;\quad-z^{\ast}=c;\quad y^{\ast}=-b;\quad w^{\ast}=a
\]
Using the equalities in Eq. \ref{invcond}, we conclude%
\begin{align*}
1  &  =ad^{\ast}-bc^{\ast}\\
0  &  =cd^{\ast}-dc^{\ast}=-ab^{\ast}+ba^{\ast}%
\end{align*}
and interchanging $g(a,b,c,d)$ with $g(a,b,c,d)^{-1}$ that%
\begin{align*}
1  &  =a^{\ast}d-c^{\ast}b\text{ }\\
0  &  =-c^{\ast}a+a^{\ast}c=-d^{\ast}b+b^{\ast}d.
\end{align*}
One may show that the set of such $g(a,b,c,d)$ compose a group under matrix
multiplication. \ Denote this group SL$_{\ast}(2,A)$. \ For $t_{x}^{p}%
t_{y}^{q}t_{z}^{\text{tr}(t)}\in\,$H$_{\ast}(A)$ we have%

\begin{equation}
g(a,b,c,d)\left(  t_{x}^{p}t_{y}^{q}t_{z}^{tr(t)}\right)  g(a,b,c,d)^{-1}%
=t_{x}^{pd^{\ast}-qc^{\ast}}t_{y}^{qa^{\ast}-pb^{\ast}}t_{z}^{\text{tr}%
(t)+pq^{\ast}-(pd^{\ast}-qc^{\ast})(qa^{\ast}-pb^{\ast})^{\ast}}%
\end{equation}

The semidirect product SL$_{\ast}(2,A)\ltimes$H$_{\ast}(A)$ is the Jacobi
group associated with $A$ . \ Berndt and Schmidt (\cite{BS} p 3) describe the
Jacobi group when $A$ is a commutative ring with identity. \ 

When the Heisenberg group H$_{\ast}(S)$ for $S\subset A$ is constructed only
over a vector space and\ not a ring we must restrict SL$_{\ast}(2,A)$ to the
subgroup whose conjugate action preserves this Heisenberg group. \ In this
paper we focus on Weil representations for SL$(2,F_{N})$ where $F_{N}$ is the
prime field $F_{N}$ for odd prime $N$. \ \ SL$(2,F_{N})$ is an automorphism
group for each of the Heisenberg groups H$_{\ast}(S)$ that we consider. \ The
automorphisms that SL$(2,F_{N})$ induce in H$_{\ast}(A)$ are summarized in
Sect. 3.4 below.

\subsection{\bigskip Semidirect product with diagonal subgroup}

Let $A_{0}^{x}$ denote the set of norm 1 elements of A%
\[
A_{0}^{x}=\{x\in A\mid xx^{\ast}=1\}.
\]
When $A$ is a Clifford algebra, $xx^{\ast}$ is the spinor norm and $A_{0}^{x}$
composes the spin group \cite{BT}. \ For $r,s,v,w\in A_{0}^{x}$ consider the
similarity transformation%
\begin{align*}
&  \left(
\begin{array}
[c]{cccc}%
r &  &  & \\
& s &  & \\
&  & v & \\
&  &  & w
\end{array}
\right)  \left(
\begin{array}
[c]{cccc}%
1 & 0 & 0 & q^{\ast}\\
p & 1 & q & \text{tr}(t)+pq^{\ast}\\
0 & 0 & 1 & -p^{\ast}\\
0 & 0 & 0 & 1
\end{array}
\right)  \left(
\begin{array}
[c]{cccc}%
r^{-1} &  &  & \\
& s^{-1} &  & \\
&  & v^{-1} & \\
&  &  & w^{-1}%
\end{array}
\right) \\
&  =\left(
\begin{array}
[c]{cccc}%
1 & 0 & 0 & rq^{\ast}w^{-1}\\
spr^{-1} & 1 & sqv^{-1} & s\left(  \text{tr}(t)+pq^{\ast}\right)  w^{-1}\\
0 & 0 & 1 & -vp^{\ast}w^{-1}\\
0 & 0 & 0 & 1
\end{array}
\right)
\end{align*}
\ We require that the central subgroup of H$_{\ast}(A)$also commute with this
transformation.\ \ Since for central involutions tr$(t)$ is in the center of
$A$, we obtain the condition $s=w.$ \ Closure of H$_{\ast}(A)$ under this
similarity transformation also requires $\left(  spr^{-1}\right)  ^{\ast
}=vp^{\ast}w^{-1}$ and rearranging $p^{\ast}=r^{-1}vp^{\ast}w^{-1}s$ for all
$p.$ \ A similar relation holds for $q.$ Using $s=w$ we obtain
\[
p=p\left(  rv^{\ast}\right)
\]
For invertible $p,$ this implies $r=v.$ \ 

We conclude that H$_{\ast}(A)$ is closed under similarity transformations with%
\[
t_{A}^{r,s}=^{\text{def}}\left(
\begin{array}
[c]{cccc}%
r &  &  & \\
& s &  & \\
&  & r & \\
&  &  & s
\end{array}
\right)  ;r,s\in A_{0}^{x}%
\]
Let D$(A_{0}^{x})$ denote the group that is composed of such matrices with
their multiplication and form the semidirect product D$(A_{0}^{x})\ltimes
$H$_{\ast}(A).$ \ The automorphisms that D$(A_{0}^{x})$ induces in H$_{\ast
}(A)$ are summarized in Sect. 3.4 below.

When considering Heisenberg groups H$_{\ast}(S)$ for $S$ a vector subspace of
an algebra, we restrict $t_{A}^{r,s}$ to the subgroup of D$(A_{0}^{x})$ that
preserves H$_{\ast}(S)$ under conjugation. \ Consider, for example, when $S$
corresponds to the subspace of $A$ with elements $x+x^{\ast}$. \ The
Heisenberg group over this subspace has general element%
\[
\left(  t_{x}^{p+p^{\ast}}t_{y}^{q+q^{\ast}}t_{z}^{\text{tr}(t)}\right)  .
\]
Conjugating with $t_{A}^{r,s}\in$D$(A_{0}^{x})$ leads to%
\[
t_{A}^{r,s}\left(  t_{x}^{p+p^{\ast}}t_{y}^{q+q^{\ast}}t_{z}^{\text{tr}%
(t)}\right)  t_{A}^{r^{\ast},s^{\ast}}=t_{x}^{s(p+p^{\ast})r^{\ast}}%
t_{y}^{s(q+q^{\ast})r^{\ast}}t_{z}^{\text{tr}(t)}%
\]
H$_{\ast}(S)$ is stable under this transformation if we restrict to $s=r.$

\subsection{Summary of group relations}

\bigskip The Heisenberg group H$_{\ast}(A)$ is composed of elements%
\[
\{t_{x}^{p}t_{y}^{q}t_{z}^{\text{tr}(t)}=\left(
\begin{array}
[c]{cccc}%
1 & 0 & 0 & q^{\ast}\\
p & 1 & q & \text{tr}(t)+pq^{\ast}\\
0 & 0 & 1 & -p^{\ast}\\
0 & 0 & 0 & 1
\end{array}
\right)  \mid p,q,t\in A\}
\]
with group multiplication%
\[
\left(  t_{x}^{p}t_{y}^{q}t_{z}^{\text{tr}(t)}\right)  \left(  t_{x}%
^{p^{\prime}}t_{y}^{q^{\prime}}t_{z}^{\text{tr}(t^{\prime})}\right)
=t_{x}^{p+p^{\prime}}t_{y}^{q+q^{\prime}}t_{z}^{\text{tr}(t+t^{\prime
}-qp^{\prime\ast})}%
\]
and commutation relation%
\[
t_{x}^{p}t_{y}^{q}=t_{y}^{q}t_{x}^{p}t_{z}^{\text{tr}(qp^{\ast})}%
\]

The SL$(2,F_{N})$ subgroup is composed of the set of matrices%
\[
\{g=\left(
\begin{array}
[c]{cccc}%
a & 0 & b & 0\\
0 & 1 & 0 & 0\\
c & 0 & d & 0\\
0 & 0 & 0 & 1
\end{array}
\right)  \mid a,b,c,d\in F_{N};\det g=1\}.
\]
It induces the following automorphisms of H$(A)$%
\begin{align}
t_{s}^{a}\left(  t_{x}^{p}t_{y}^{q}t_{z}^{\text{tr}(t)}\right)  t_{s}%
^{\frac{1}{a}}  &  =t_{x}^{\frac{p}{a}}t_{y}^{qa}t_{z}^{\text{tr}%
(t)}\nonumber\\
t_{d}^{c}\left(  t_{x}^{p}t_{y}^{q}t_{z}^{\text{tr}(t)}\right)  t_{d}^{-c}  &
=t_{x}^{(p-qc)}t_{y}^{q}t_{z}^{\text{tr}(t)+cqq^{\ast}}\nonumber\\
t_{u}^{b}\left(  t_{x}^{p}t_{y}^{q}t_{z}^{\text{tr}(t)}\right)  t_{u}^{-b}  &
=t_{x}^{p}t_{y}^{q-bp}t_{z}^{\text{tr}(t)+bpp^{\ast}}\nonumber\\
j\left(  t_{x}^{p}t_{y}^{q}t_{z}^{\text{tr}(t)}\right)  j^{-1}  &  =t_{x}%
^{q}t_{y}^{-p}t_{z}^{\text{tr}(t+pq^{\ast})}\nonumber\\
t_{r}^{a+\sqrt{\delta}b}t_{x}^{p}t_{y}^{q}t_{z}^{\text{tr}(t)}t_{r}%
^{a-\sqrt{\delta}b}  &  =t_{x}^{pa-qb}t_{y}^{-pb\delta+qa}t_{z}^{\text{tr}%
(t)+pq^{\ast}-(pa-qb)(-pb\delta+qa)^{\ast}} \label{sl2 action}%
\end{align}
The SL$(2,F_{N})$ operators in these expressions are defined in section IV.A below.

The diagonal subgroup D$(A_{0}^{x})$ is composed of matrices
\[
\{t_{A}^{r,s}=\left(
\begin{array}
[c]{cccc}%
r &  &  & \\
& s &  & \\
&  & r & \\
&  &  & s
\end{array}
\right)  ;r,s\in A_{0}^{x}\}
\]
and generates the following transformation of H$_{\ast}(A)$%
\[
t_{A}^{r,s}\left(  t_{x}^{p}t_{y}^{q}t_{z}^{\text{tr}(t)}\right)
t_{A}^{r^{-1},s^{-1}}=t_{x}^{spr^{-1}}t_{y}^{sqr^{-1}}t_{z}^{\text{tr}(t)}%
\]
The action of D$(A_{0}^{x})$ preserves the norm and acts on both sides of the
arguments of the translation subgroups of the Heisenberg group. $\ $%
SL$(2,F_{N})$ commutes with D$(A_{0}^{x})$.

\subsection{Example: \ $C4\otimes Q$ group and algebra}

Since our motivation derives primarily from physics and since the complex
quaternion algebra suffices to treat cases of immediate interest let us focus
on this algebra. \ The complex quaternion algebra is the algebra of 2x2
complex matrices. \ It is also known as the Pauli algebra. \ \ It is the
Clifford algebra $C_{3,0}$ in the notation of Benn and Tucker \cite{BT}. \ It
is considered in a different context in \cite{J1}.

The complex quaternion group $C4\otimes Q$ has generators%
\[
\{e_{1},e_{2},e_{3}\}
\]
and relations%
\[
e_{1}^{2}=e_{2}^{2}=e_{3}^{2}=1\text{ };(e_{ij})^{4}=e_{123}^{4}=1\text{ for
}i\neq j
\]
where we use the notation $e_{i}e_{j}=e_{ij}$ $.$ \ So that we may construct
group representations in an elementary way, we consider the complex quaternion
algebra over an odd prime field $F_{N}.$ \ We restrict to the subspace whose
representation maps $(e_{ij})^{2}=-1$ for $i\neq j$. \ We then also have
$\left(  e_{123}\right)  ^{2}=-1.$ \ $C4\otimes Q$ is isomorphic with
$C4\otimes$D$4$ where $D4$ is the dihedral group with order 4$.$ \ The center
of the group is $\{1,e_{123}\}.$

Let us denote the group algebra for $C4\otimes Q$ over the prime field $F_{N}$
as $C4\otimes Q(F_{N}).$ \ A general element in this algebra can be written%
\begin{align}
x  &  =x_{0}1+x_{1}e_{1}+x_{2}e_{2}+x_{3}e_{3}+x_{4}e_{12}+x_{5}e_{23}%
+x_{6}e_{31}+x_{7}e_{123}\nonumber\\
&  \label{element}%
\end{align}
We may express the basis elements of $C4\otimes Q$ as 2x2 matrices%
\begin{align*}
1  &  =\left(
\begin{array}
[c]{cc}%
1 & 0\\
0 & 1
\end{array}
\right)  ,e_{1}=\left(
\begin{array}
[c]{cc}%
0 & 1\\
1 & 0
\end{array}
\right)  ,e_{2}=\left(
\begin{array}
[c]{cc}%
0 & -i\\
i & 0
\end{array}
\right)  ,e_{3}=\left(
\begin{array}
[c]{cc}%
1 & 0\\
0 & -1
\end{array}
\right)  ,\\
e_{12}  &  =\left(
\begin{array}
[c]{cc}%
i & 0\\
0 & -i
\end{array}
\right)  ,e_{23}=\left(
\begin{array}
[c]{cc}%
0 & i\\
i & 0
\end{array}
\right)  ,e_{31}=\left(
\begin{array}
[c]{cc}%
0 & 1\\
-1 & 0
\end{array}
\right)  ,e_{123}=\left(
\begin{array}
[c]{cc}%
i & 0\\
0 & i
\end{array}
\right)
\end{align*}
The general element $x$ in the group algebra of $C4\otimes Q$ can then be
written in matrix form as
\[
x=\left(
\begin{array}
[c]{cc}%
x_{0}+x_{3}+ix_{4}+ix_{7} & \quad x_{1}-ix_{2}+ix_{5}+x_{6}\\
x_{1}+ix_{2}+ix_{5}-x_{6} & \quad x_{0}-x_{3}-ix_{4}+ix_{7}%
\end{array}
\right)
\]
We consider the following maps%
\begin{align*}
x^{\eta}  &  =x_{0}-x_{1}e_{1}-x_{2}e_{2}-x_{3}e_{3}+x_{4}e_{12}+x_{5}%
e_{23}+x_{6}e_{31}-x_{7}e_{123}\\
x^{\xi}  &  =x_{0}+x_{1}e_{1}+x_{2}e_{2}+x_{3}e_{3}-x_{4}e_{12}-x_{5}%
e_{23}-x_{6}e_{31}-x_{7}e_{123}\\
x^{\xi\eta}  &  =x_{0}-x_{1}e_{1}-x_{2}e_{2}-x_{3}e_{3}-x_{4}e_{12}%
-x_{5}e_{23}-x_{6}e_{31}+x_{7}e_{123}\\
&  =^{\text{def}}x^{\ast}%
\end{align*}
The map $\eta$ is the main automorphism and is obtained by changing the signs
of each of the group generators $\{e_{1},e_{2},e_{3}\}.$ \ The reversal
mapping $\xi$ is the anti-automorphism obtained by taking each group element
to its inverse. \ It corresponds to the matrix conjugate transpose (or
adjoint) mapping when $x$ is given in matrix form. \ \ We compose $\eta$ and
$\xi$ to obtain the central involution $\xi\eta$ that plays the role of the
central involution $\ast$.

We have the trace mapping
\begin{align*}
\text{tr}(xy)  &  =(xy)+(xy)^{\ast}\\
&  =2[(x_{0}y_{0}+x_{1}y_{1}+x_{2}y_{2}+x_{3}y_{3}-x_{4}y_{4}-x_{5}y_{5}%
-x_{6}y_{6}-x_{7}y_{7})1\\
&  +(x_{0}y_{7}+x_{7}y_{0}+x_{1}y_{5}+x_{5}y_{1}+x_{2}y_{6}+x_{6}y_{2}%
+x_{3}y_{4}+x_{4}y_{3})e_{123}]
\end{align*}
and observe
\[
\text{tr}(xy)=\text{tr}(yx)
\]
We have the norm mapping \
\begin{align*}
n(x)  &  =xx^{\ast}\\
&  =\left(  x_{0}^{2}-x_{1}^{2}-x_{2}^{2}-x_{3}^{2}+x_{4}^{2}+x_{5}^{2}%
+x_{6}^{2}-x_{7}^{2}\right)  1+2(x_{0}x_{7}-x_{1}x_{5}-x_{2}x_{6}-x_{3}%
x_{4})e_{123}%
\end{align*}
with product rule
\[
n(xy)=xyy^{\ast}x^{\ast}=n(x)n(y),
\]
since $yy^{\ast}$ is in the center of the algebra. \ The norm $n(x)$ of an
element $x\in C4\otimes Q(F_{N})$ corresponds to the determinant of its matrix form.

The identity component of the trace can be expressed%
\begin{align*}
\text{Tr}(x)  &  =\frac{1}{2}(x+x^{\xi\eta}+x^{\eta}+x^{\xi})\\
&  =2x_{0}%
\end{align*}
The identity component of the norm can be expressed%
\begin{align*}
N(x)  &  =\frac{1}{2}(n(x)+n(x)^{\xi})\\
&  =\left(  x_{0}^{2}-x_{1}^{2}-x_{2}^{2}-x_{3}^{2}+x_{4}^{2}+x_{5}^{2}%
+x_{6}^{2}-x_{7}^{2}\right)
\end{align*}

The $C4\otimes Q(F_{N})$ algebra with norm $n(x)=xx^{\ast}$ can be viewed as a
4d complex space with Euclidean norm. \ Here $e_{123}$ plays the role of the
unit imaginary number. \ Subspaces of this algebra provide finite models for
each of the real or complex quadratic spaces with dimension 4 or less.

Let us consider the case of a real vector space with signature (1,3) in more
detail. \ The subspace $S$ $\subset C4\otimes Q(F_{N})$ that contains
elements
\[
x+x^{\xi}=2(x_{0}1+x_{1}e_{1}+x_{2}e_{2}+x_{3}e_{3})
\]
provides an example of such a vector space. \ We have the norm
\[
n(x+x^{\xi})=4\left(  x_{0}^{2}-x_{1}^{2}-x_{2}^{2}-x_{3}^{2}\right)  .
\]
and bilinear form
\[
\text{tr}((x+x^{\xi})(y+y^{\xi})^{\xi\eta})=8(x_{0}y_{0}-x_{1}y_{1}-x_{2}%
y_{2}-x_{3}y_{3})
\]
We form the Heisenberg group over this subspace having elements $\{t_{x}%
^{p+p^{\xi}}t_{y}^{q+q^{\xi}}t_{z}^{\text{tr}(t)}\mid p,q\in C4\otimes
Q(F_{N}),$tr$(t)\in F_{N}\}.$ \ \ This Heisenberg group is closed under the
norm preserving automorphisms%
\[
t_{A}^{r,r^{\eta}}\left(  t_{x}^{p+p^{\xi}}t_{y}^{q+q^{\xi}}t_{z}%
^{\text{tr}(t)}\right)  t_{A}^{r^{\xi\eta},r^{\xi}}=t_{x}^{r^{\eta}(p+p^{\xi
})r^{\xi\eta}}t_{y}^{r^{\eta}(q+q^{\xi\eta})r^{\xi\eta}}t_{z}^{\text{tr}(t)}%
\]
The subgroup $\{t_{A}^{r,r^{\eta}}\mid r\in C4\otimes Q(F_{N});rr^{\xi\eta
}=1\}\subset$D$(A_{0}^{x})$ is isomorphic to SL$(2,C(F_{N}))$ that is our
finite counterpart to SL$(2,C),$ the two fold cover of the proper
orthochronous Lorentz group$.$ The group formed by the semidirect product of
SL$(2,C(F_{N}))$ with the translation subgroup T$_{x}=\{t_{x}^{p+p^{\xi}}\mid
p\in C4\otimes Q(F_{N})\}$ is the Poincare group. \ \ In this way we place the
Poincare group within the larger group structure $\left(  \text{SL}%
(2,F_{N})\otimes\text{D}(A_{0}^{x})\right)  \ltimes$H$_{\ast}(S).$

\bigskip Tables 1 and 2 show \ basis vectors in $C4\otimes Q$ that span
subspaces with representative signatures. \ In these tables $C2=\{1,e_{1}\}$
denotes the cyclic group with order 2, \ $C4=\{1,e_{123}\}$ in Table 1 and
$C4=\{1,e_{12}\}$ in Table 2 denotes the cyclic group with order 4,
$Q=\{1,e_{12},e_{23,}e_{31}\}$ denotes the quaternion group, and
$D4=\{1,e_{1},e_{2},e_{12}\}$ is the dihedral group of order 4.

\pagebreak

Table 1. \ \ Complex vector spaces. \ $e_{123}$ plays role of unit imaginary number.\ %

\begin{tabular}
[c]{ccc}\hline\hline
basis elements & subspace & signature\\\hline
$1,e_{123}$ & $C4(F_{N})$ & (1,0)\\
$1,e_{12},e_{3,}e_{123}$ & $C2\otimes C4(F_{N})$ & (2,0)\\
$e_{1},e_{2},e_{3},e_{12},e_{23},e_{31}$ & $x-x^{\xi\eta}\subset C4\otimes
Q(F_{N})$ & (3,0)\\
$1,e_{1},e_{2},e_{3},e_{12},e_{23},e_{31},e_{123}$ & $C4\otimes Q(F_{N})$ &
(4,0)
\end{tabular}

\bigskip

Table 2. \ Vector spaces over $F_{N}.$%

\begin{tabular}
[c]{ccc}\hline\hline
basis elements & subspace & signature\\\hline
$1,e_{12}$ & $C4(F_{N})$ & (2,0)\\
$1,e_{1}$ & $C2(F_{N})$ & (1,1)\\
$e_{12},e_{23},e_{31}$ & $x-x^{\xi}\subset Q(F_{N})$ & (3,0)\\
$1,e_{1},e_{2}$ & $x+x^{\xi}\subset D4(F_{N})$ & (1,2)\\
$1,e_{12},e_{23,}e_{31}$ & $Q(F_{N})$ & (4,0)\\
$1,e_{1},e_{2},e_{12}$ & $D4(F_{N})$ & (2,2)\\
$1,e_{1},e_{2},e_{3}$ & $x+x^{\xi}\subset C4\otimes Q(F_{N})$ & (1,3)
\end{tabular}

\section{Group Representations}

In this section we construct a model for the representations of $($%
D$(A_{0}^{x})\otimes$SL$(2,F))\ltimes$H$_{\ast}(S)$ when $S\subset$ $C4\otimes
Q(F_{N}),$ $F$ is the field $F_{N}$ for odd prime $N$ and $A_{0}^{x}$ is the
group of unit norm elements in the algebra containing $S$. \ In order that our
construction be self-contained and to define notation let us start with the
representations of important cyclic subgroups.

\subsection{Representations of the cyclic subgroups}

\bigskip Let us describe a \ model for representations of the abelian
translation subgroup
\begin{align*}
\text{T}_{x}  &  =\{t_{x}^{h}\mid h\in S\}\\
&  \text{with multiplication}\\
t_{x}^{h}t_{x}^{h^{\prime}}  &  =t_{x}^{h+h^{\prime}}%
\end{align*}
We form eigenstates%
\begin{align}
\widehat{x}_{\eta}  &  =\frac{1}{V}%
{\displaystyle\sum\limits_{h\in S}}
\exp(\frac{-2\pi i}{N}\text{Tr}\eta h)t_{x}^{h}\nonumber\\
&  \label{estate for tx}%
\end{align}
where $V$ is the number of elements in the translation subgroup (For
$S=C4\otimes Q(F_{N})$, $V=N^{8}.)$ and $\eta\in S.$ \ Note that Tr$(\eta h)$
projects to the identity component of the center of $S$ (in contrast to
$tr(\eta h)$ that projects to the center of $S$ and that may be
multidimensional) $.$ \ We have the group action and relationships%

\begin{align*}
t_{x}^{h}\widehat{x}_{\eta}  &  =\exp(\frac{2\pi i}{N}\text{Tr}\eta
h)\widehat{x}_{\eta}\\%
{\displaystyle\sum\limits_{\eta\in S}}
\widehat{x}_{\eta}  &  =1\\
\widehat{x}_{\eta}\widehat{x}_{\eta^{\prime}}  &  =\delta(\eta-\eta^{\prime
})\widehat{x}_{\eta}%
\end{align*}
In the same way we form eigenstates $\widehat{y}_{\eta}$ of $t_{y}^{h}$ and
eigenstates $\widehat{z}_{\omega}$ of $t_{z}^{h}$ for $h,\eta,\omega\in S$. \ 

\bigskip Eigenstates for the additive subgroups of SL$(2,F_{N})$ can be
constructed in the same way. \ Writing elements of SL$(2,F_{N})$ as 2x2
matrices, we introduce notation in the following. \ For the subgroup%
\[
\ \{t_{u}^{b}=^{\text{def}}\left(
\begin{array}
[c]{cc}%
1 & b\\
0 & 1
\end{array}
\right)  \mid b\in F_{N}\}
\]
we form eigenstates%
\begin{equation}
\widehat{u}_{\sigma}=\frac{1}{N}%
{\displaystyle\sum\limits_{b\in F_{N}}}
\exp(\frac{-2\pi i}{N}\sigma b)t_{u}^{b} \label{u def}%
\end{equation}
Denote the order four element%
\[
j=^{\text{def}}\left(
\begin{array}
[c]{cc}%
0 & 1\\
-1 & 0
\end{array}
\right)
\]
For the subgroup%
\[
\{t_{d}^{c}=^{\text{def}}\left(
\begin{array}
[c]{cc}%
1 & 0\\
c & 1
\end{array}
\right)  =jt_{u}^{-c}j^{-1}\mid c\in F_{N}\}
\]
we form eigenstates
\begin{equation}
\widehat{d}_{\sigma}=\frac{1}{N}%
{\displaystyle\sum\limits_{c\in F_{N}}}
\exp(\frac{-2\pi i}{N}\sigma c)t_{d}^{c} \label{deigst}%
\end{equation}

Eigenstates for the multiplicative subgroups of SL$(2,F_{N})$ are constructed
in a similar way. \ For the multiplicative group%
\[
\ \{t_{s}^{a}=^{\text{def}}\left(
\begin{array}
[c]{cc}%
a & 0\\
0 & \frac{1}{a}%
\end{array}
\right)  \mid a=z^{(N+1)t}\in F_{N}^{x}\}
\]
we form eigenstates%
\begin{equation}
\widehat{s}_{\chi}=\frac{1}{N-1}%
{\displaystyle\sum\limits_{z^{(N+1)\chi}\in F_{N}^{x}}}
\exp(\frac{-2\pi i}{N-1}\chi t)t_{s}^{z^{(N+1)t}} \label{s def}%
\end{equation}
for $z^{(N+1)\chi}$ $\in F_{N}^{x}.$ \ Here, $z$ denotes a generator of the
multiplicative group $F_{N^{2}}^{x}$ in the quadratic extension $F_{N^{2}}.$
\ It has order $N^{2}-1$. The number $z^{(N+1)}$ is a generator of the
multiplicative group $F_{N}^{x}$ of $F_{N}$ with order $N-1.$ \ We will often
denote the action of $t_{s}^{a}$ on $\widehat{s}_{\chi}$ as $\chi(a)$ for
$\chi(a)=\exp(\frac{2\pi i}{N-1}\chi t)$ $.$ We denote the representation
having $\chi=0$ as $\widehat{s}_{+}$ and the representation with $\chi
=\frac{N-1}{2}$ as $\widehat{s}_{-}$, where
\[
\widehat{s}_{-}=\frac{1}{N-1}%
{\displaystyle\sum\limits_{a\in F_{N}^{x}}}
\left(  \frac{a}{N}\right)  t_{s}^{a}%
\]
and $\left(  \frac{a}{N}\right)  $ is the Legendre symbol.

The circle subgroup has elements
\[
\{t_{r}^{a+\sqrt{\delta}b}=^{\text{def}}\left(
\begin{array}
[c]{cc}%
a & b\delta\\
b & a
\end{array}
\right)  \}
\]
where $a+\sqrt{\delta}b=z^{(N-1)n}$ and $z^{(N-1)}$ is a generator of the
multiplicative group of unit norm elements $U$ in the quadratic extension
$F_{N^{2}}$(see, e.g., \cite{T} p 306)$.$ \ $U$ has order $N+1.$ \ We form
eigenstates%
\begin{equation}
\widehat{r}_{\mu}=\frac{1}{N+1}%
{\displaystyle\sum\limits_{a+\sqrt{\delta}b=z^{(N-1)m}\in U}}
\exp(\frac{-2\pi i}{N+1}\mu m)t_{r}^{z^{(N-1)m}} \label{tr eigenstate}%
\end{equation}
for each $z^{(N-1)\mu}\in U$. \ \ \ 

Additional information about SL$(2,F_{N})$ $\subset$GL$(2,F_{N})$ and its
subgroups and representations are contained in Terras (\cite{T}, Ch. 21) and
Piatetski-Shapiro \cite{P}.

\subsection{\bigskip Representations of the Heisenberg group}

\bigskip Let us develop models for the representations of H$_{\ast}(S).$ \ The
representatations with trivial central character can be understood very simply
as arising from the group action on ideals in its group algebra with form
$\widehat{x}_{\sigma}\widehat{y}_{\eta}\widehat{z}_{0}$ for $\sigma,\eta\in
S.$ \ $\widehat{z}_{0}$ corresponds to the trivial representation of the
central subgroup. \ The 1d subspaces $\widehat{x}_{\sigma}\widehat{y}_{\eta
}\widehat{z}_{0}$ are each stable under H$_{\ast}(S).$ \ 

\ Let us construct a model for the representations with nontrivial central
character. \ Consider the maximal abelian subgroup $\{t_{y}^{s}t_{z}%
^{\text{tr}(t)}\mid s,t\in S\}.$ and form the 1 dimensional invariant subspace
in its group algebra
\[
\widehat{y}_{\eta}\widehat{z}_{\text{tr}\omega}\text{ for tr}\omega\neq0
\]
For fixed tr$(\omega)$ any choice for $\widehat{y}_{\eta}$ leads to an
equivalent representation; for simplicity, let us choose $\eta=0.$ \ \ Act on
$\widehat{y}_{0}\widehat{z}_{tr(\omega)}$with a general element $f$ in the
H$_{\ast}(S)$ group algebra
\[
f=%
{\displaystyle\sum\limits_{h,k,l\in S}}
f(h,k,\text{tr}(l))t_{x}^{h}t_{y}^{k}t_{z}^{tr(l)}.
\]
where\ \ $f(h,k,$tr$(l))$ is a complex number for each $h,k,l\in S$ . \ We
obtain \
\begin{align*}
f\widehat{y}_{0}\widehat{z}_{tr(\omega)}  &  =\left(
{\displaystyle\sum\limits_{h,k,l\in S}}
f(h,k,\text{tr}(l))t_{x}^{h}t_{y}^{k}t_{z}^{\text{tr}(l)}\right)  \widehat
{y}_{0}\widehat{z}_{tr(\omega)}\\
&  =%
{\displaystyle\sum\limits_{h\in S}}
f(h)t_{x}^{h}\widehat{y}_{0}\widehat{z}_{\text{tr}(\omega)}%
\end{align*}
after absorbing $t_{y}^{k}t_{z}^{\text{tr}(l)}$ into $\widehat{y}_{0}%
\widehat{z}_{\omega-\varepsilon\omega}$ and summing over $k,l$. \ Here $f(h)$
a complex number for each $h.$ Let us denote $f\widehat{y}_{0}\widehat
{z}_{\text{tr}(\omega)}=f_{\text{tr}(\omega)}.$ \ The action of the Heisenberg
group on $f_{\text{tr}(\omega)}$ is the Schr\"{o}dinger representation with
central character tr$(\omega).$ \ \ The Schr\"{o}dinger representation is
unique and irreducible for each value of \ tr$(\omega)\neq0$. \ It has
dimensionality $V$ the number of elements in $S.$ \ See, Terras (\cite{T}, ch
18) for additional description of the Heisenberg group and its representations.

Similarity transformation with
\[
M=\left(
\begin{array}
[c]{cccc}%
x &  &  & \\
& y &  & \\
&  & \frac{yw}{x} & \\
&  &  & w
\end{array}
\right)  \text{for }x,y,w\in\text{center of A and nonzero}%
\]
leads to
\[
M\left(  t_{x}^{p}t_{y}^{q}t_{z}^{\text{tr}(t)}\right)  M^{-1}=t_{x}^{\frac
{y}{x}p}t_{y}^{\frac{x}{w}q}t_{z}^{\frac{y}{w}\text{tr}(t)}%
\]
This transformation can be used to translate between Schr\"{o}dinger
representations having different central character. \ In the following, we
treat the special case when the central character tr$(\omega)=\omega_{0}$ is
in the ground field $F_{N}.$ \ \ The action of H$_{\ast}(S)$ on $f_{\omega
_{0}}$ is summarized below in Eqs. $\ref{action of ty}$ .

\subsection{A representation of \bigskip SL$(2,F_{N})\ltimes$H$_{\ast}(S)$}

We now describe a model for the Schr\"{o}dinger-Weil representation of the
semidirect product group \bigskip SL$(2,F_{N})\ltimes$H$_{\ast}(S)$ for $S$ a
subspace of $C4\otimes Q(F_{N})$. \ This construction is similar to the case
when $A$ is 1 dimensional \cite{J2}. \ The Weil representation for the group
SL$_{\ast}(2,A)$ that contains SL$(2,F_{N})$ is described by Gutierrez et. al.
\cite{Gu1}.

We consider a subspace of the regular representation of SL$(2,F_{N})\ltimes
$H$_{\ast}(S)$ having vectors with the form%
\begin{equation}
f=%
{\displaystyle\sum\limits_{k\in S}}
f(k)t_{x}^{k}I \label{swrep1}%
\end{equation}
for
\begin{align}
I  &  =\widehat{y}_{0}\widehat{z}_{\omega_{0}}\widehat{s}_{\chi}\widehat
{u}_{0}\left\{
{\displaystyle\sum\limits_{p\in S}}
t_{x}^{p}\left(  1+\tau%
{\displaystyle\sum\limits_{l\in S;N(l)=0}}
\exp(\frac{2\pi i}{N}2\omega_{0}\text{Tr}lp^{\ast})\right)  \right\}
(1+\alpha j)\widehat{d}_{0}\nonumber\\
&  \label{ideal}%
\end{align}
In this expression $\tau,\alpha$ are numbers to be determined, $\widehat
{s}_{\chi}$ is either $\widehat{s}_{+}$ or $\widehat{s}_{-}$ and we restrict
to central characters $\omega_{0}\in F_{N}^{x}$. \ \ We show that $f$ resides
in a stable subspace under the action of SL$(2,F_{N})\ltimes$H$_{\ast}%
(S)$\ for appropriate choice of the numbers $\tau,\alpha$ and choice of
$\widehat{s}_{\chi}$. The action of SL$(2,F_{N})\ltimes$H$_{\ast}(S)$ on $f$
then determines the sought after representation.

It is evident by inspection that $f$ is stable under H$(S).$ \ One can also
show without difficulty that $f$ is stable under the action of $t_{u}^{b}\in
$SL$(2,F_{N}),b\in F_{N}.$ \ This is done by expanding out the terms in Eq.
\ref{swrep1} to the left of $\widehat{u}_{0}$, passing $t_{u}^{b}$ through
these terms using the group relations, and then absorbing $t_{u}^{b}$ into
$\widehat{u}_{0}.$ \ Since SL$(2,F_{N})$ can be generated using the 2
operators $t_{u}^{1}$ and $j$ it is left to show that $f$ is stable under the
action of $j.$ \ We verify this by a direct calculation that is summarized in
Appendix A.

\subsubsection{Group actions in the Schr\"{o}dinger-Weil representation}

We have the following group actions for the Heisenberg group generators%
\begin{align}
t_{x}^{h}f  &  =%
{\displaystyle\sum\limits_{k\in S}}
f(k-h)t_{x}^{k}I\label{action of tx}\\
t_{y}^{h}f  &  =%
{\displaystyle\sum\limits_{k\in S}}
\exp(\frac{-2\pi i}{N}\omega_{0}2\text{Tr}kh^{\ast})f(k)t_{x}^{k}%
I\label{action of ty}\\
t_{z}^{l}f  &  =\exp(\frac{2\pi i}{N}\omega_{0}\text{Tr}l)f\nonumber
\end{align}
For the SL$(2,N)$ group operators we have
\begin{align}
t_{u}^{b}f  &  =%
{\displaystyle\sum\limits_{k\in S}}
\exp(\frac{2\pi i}{N}\omega_{0}b\text{Tr}n(k))f(k)t_{x}^{k}I \label{tu action}%
\\
t_{s}^{a}f  &  =\left(  \frac{a}{N}\right)  ^{\dim}%
{\displaystyle\sum\limits_{k\in S}}
f(ak)t_{x}^{k}I\text{ }\label{ts action}\\
jf  &  =\kappa%
{\displaystyle\sum\limits_{h\in S}}
\left(
{\displaystyle\sum\limits_{k\in S}}
\exp\left(  \frac{2\pi i}{N}\omega_{0}2\text{Tr}kh^{\ast}\right)  f(k)\right)
t_{x}^{h}I
\end{align}
where%
\[
\kappa=\frac{1}{V}G(1,N)^{\dim}\left(  \frac{2\omega_{0}}{N}\right)  ^{\dim
}\left(  \frac{-1}{N}\right)  ^{q}%
\]
In this expression, dim is the dimensionality of $S$ and $q$ is the number of
negative terms in the quadratic form. \ The Gauss sum $G(1,N)$ (\cite{L}
p\ 82) is given by $\ $%
\begin{align*}
G(1,N)  &  =%
{\displaystyle\sum\limits_{a\in F_{N}^{x}}}
\left(  \frac{a}{N}\right)  \exp(\frac{2\pi i}{N}a)\\
&  =%
{\displaystyle\sum\limits_{a\in F_{N}}}
\exp(\frac{2\pi i}{N}a^{2})\\
&  =\{_{i\sqrt{N}\text{ for }N=3\operatorname{mod}4}^{\sqrt{N}\text{ for
}N=1\operatorname{mod}4}%
\end{align*}
To calculate $t_{\ d}^{c}f$ we use%
\[
t_{\ d}^{c}f=jt_{u}^{-c}j^{-1}f
\]
and substitute into this expression the action of the operators $j,$
$t_{u}^{-c}$ and $j^{-1}=jt_{s}^{-1}.$ \ We obtain%
\begin{align}
t_{\ d}^{c}f  &  =\frac{1}{V}%
{\displaystyle\sum\limits_{l\in S}}
\left(
{\displaystyle\sum\limits_{h,k\in S}}
\exp(\frac{2\pi i}{N}\omega_{0}\text{Tr}(-cn(h)+2h^{\ast}(k-l)))f(k)\right)
t_{x}^{l}I\nonumber\\
&  \label{td action}%
\end{align}
We may obtain $t_{r}^{a+\sqrt{\delta}b}$ by using either of the
decompositions
\begin{align*}
t_{r}^{a+\sqrt{\delta}b}  &  =t_{d}^{b/a}t_{s}^{a}t_{u}^{b\delta/a}\\
&  =t_{u}^{\frac{a-1}{b}}t_{d}^{b}t_{u}^{\frac{a-1}{b}}%
\end{align*}
We find for $b\neq0$%
\begin{align}
&  t_{r}^{a+\sqrt{\delta}b}f\nonumber\\
&  =\frac{1}{V}%
{\displaystyle\sum\limits_{h\in A}}
{\displaystyle\sum\limits_{p,k\in A}}
\exp\left(  \frac{2\pi i}{N}\omega_{0}\text{Tr}((n(k)+n(h))(\frac{a-1}%
{b})-\frac{b}{4\omega_{0}^{2}}n(p))\right)  \exp\left(  \frac{2\pi i}%
{N}\text{Tr}p(k-h)\right)  f(k)t_{x}^{h}I\nonumber\\
&  \label{acttr}%
\end{align}
\ The actions for \ $t_{\ d}^{c}$ and $t_{r}^{a+\sqrt{\delta}b}$ can be
reexpressed by completing the norm and summing over $h$ in Eq. \ref{td action}
\ and over $p$ in Eq. \ref{acttr}. \ The given form for $t_{\ d}^{c}$ is
convenient for use below.

\subsection{\bigskip Representations including the diagonal subgroup}

To obtain representations of the complete group $($SL$(2,F_{N})\otimes
$D$(A_{0}^{x}))\ltimes$H$(S)$ consider the action of $t_{A}^{r,s}\in$%
D$(A_{0}^{x})$ on the function $f=%
{\displaystyle\sum\limits_{k\in S}}
f(k)t_{x}^{k}I$
\[
t_{A}^{r,s}f=%
{\displaystyle\sum\limits_{k\in S}}
f(s^{-1}kr)t_{x}^{k}It_{A}^{r,s}%
\]
$t_{A}^{r,s}$ generates a norm-preserving transformation of the argument of
$t_{x}^{k}$ (after redefining $k$ this becomes a change in the argument of
$f(k)$) and commutes through the remainder of the expression. \ The function
$f$ will be stable under this action if we place a left ideal of D$(A_{0}%
^{x})$ on the right of $f$ to absorb $t_{A}^{r,s}.$ \ Most simply, we can
place the ideal
\[
\rho_{0}=\sum\limits_{r,s\in A_{0}^{x}}t_{A}^{r,s}%
\]
on the right hand side of $f.$ \ $\ t_{A}^{r,s}$ \ has trivial action on
$\rho_{0}.$ \ The function $f(k)$ then behaves like a scaler function of a
vector variable$.$ \ 

Now I show schematically how finite dimensional representatations of
D$(A_{0}^{x})$ arise in this construction. \ Form the affine-like group
D$(A_{0}^{x})\ltimes\,$T$_{x}$ (the Poincare group is of this type) with
elements
\[
\{t_{x}^{h}t_{A}^{r,s}\}
\]
For $\tau\in A(F_{N})$ let $O_{\tau}=\{r\tau s^{\ast}\mid r,s\in A_{0}^{x}\}.$
\ $O_{\tau}$ is the orbit of $\tau$ for this action$.$ \ The subspace in the
D$(A_{0}^{x})\ltimes\,$T$_{x}$ group algebra that contains vectors%
\[
f_{A}=%
{\displaystyle\sum\limits_{\sigma\in O_{\tau}}}
f_{A}(\sigma)\widehat{x}_{\sigma}\rho_{0},
\]
for $f_{A}(\sigma)$ a complex number for each $\sigma,$ is stable under the
group action:%
\[
t_{x}^{h}t_{A}^{r,s}f_{A}=%
{\displaystyle\sum\limits_{\sigma\in O_{\tau}}}
\exp(\frac{2\pi i}{N}\text{Tr}h\sigma)f_{A}(r^{-1}\sigma s)\widehat{x}%
_{\sigma}\rho_{0}.
\]
\ Placing $f_{A}$ on the right side of Eq. \ref{swrep1} we find a
representation for $($D$(A_{0}^{x})\otimes$SL$(2,F))\ltimes$H$_{\ast}(A)$
given by the action of the group on the vector. \
\[
f=%
{\displaystyle\sum\limits_{k\in A;\sigma\in O_{\tau}}}
f(k,\sigma)t_{x}^{k}I\widehat{x}_{\sigma}\rho_{0}%
\]
Now, express $\widehat{x}_{\sigma}=\frac{1}{V}%
{\displaystyle\sum\limits_{h\in A}}
\exp(\frac{-2\pi i}{N}$Tr$h\sigma)t_{x}^{h}$ and expand out the exponential in
this expression. \ Observe that the $m^{th}$ order term in the expansion of
the exponential corresponds to a vector in the space of homogeneous
polynomials of degree m; also, note that the space containing this vector is
stable under $t_{A}^{r,s}.$ \ These spaces of homogeneous polynomials
correspond to the representation spaces of finite dimensional representations
of D$(A_{0}^{x}).$

When D$(A_{0}^{x})\ltimes\,$T$_{x}$ corresponds to the Poincare group the
construction of the finite dimensional representations is well known\ (See,
e.g., Sternberg \cite{S} p 143.). \ \ Berndt describes the unitary (\cite{B} p
143) representations when D$(A_{0}^{x})$ is the Lorentz group. \ Examples of
these representations can be realized in the above construction by placing an
ideal in the group algebra of D$(A_{0}^{x})$ (rather than in the group algebra
of D$(A_{0}^{x})\ltimes\,$T$_{x}$) \ on the right hand side of \ Eq.
\ref{swrep1}.

\section{\bigskip Pseudo-derivative operator}

In this section we construct a pseudo-derivative operator that is applicable
to functions in the group algebra of a finite group. \ \ These operators
enable a comparison of our results to those obtained using differential
operators on functions defined over real spaces. \ 

\subsection{Construction of pseudo-derivative operator}

Act on an element $t_{x}^{h}$ $\ $in the additive cyclic group T$_{x}%
(A)=\{t_{x}^{h}\mid h\in A\}$ with $1=%
{\displaystyle\sum\limits_{\sigma\in A}}
\widehat{x}_{\sigma}$%
\begin{align}
t_{x}^{h}  &  =t_{x}^{h}%
{\displaystyle\sum\limits_{\sigma\in A}}
\widehat{x}_{\sigma}\nonumber\\
&  =%
{\displaystyle\sum\limits_{\sigma\in A}}
\exp(\frac{2\pi i}{N}\text{Tr}h\sigma)\widehat{x}_{\sigma}\nonumber
\end{align}
Since $\widehat{x}_{\sigma}^{m}=\widehat{x}_{\sigma}$we can write
\[
t_{x}^{h}=%
{\displaystyle\sum\limits_{\sigma\in A}}
{\displaystyle\sum\limits_{m=0}^{\infty}}
\frac{1}{m!}(\frac{2\pi i}{N}\text{Tr}(h\sigma)\widehat{x}_{\sigma})^{m}%
\]
and since $\widehat{x}_{\sigma}\widehat{x}_{\sigma^{\prime}}=\delta
(\sigma-\sigma^{\prime})\widehat{x}_{\sigma}$ we switch the order of the
summations and obtain
\begin{equation}
t_{x}^{h}=\exp(\frac{2\pi i}{N}\text{Tr}(h%
{\displaystyle\sum\limits_{\sigma\in A}}
\sigma\widehat{x}_{\sigma})) \label{tx lie}%
\end{equation}
The exponential map Eq. \ref{tx lie} from an element in the group algebra to
an element in the group is analogous to the exponential map between Lie
algebra and Lie group elements. \ It motivates that we consider the operator%
\[
X=%
{\displaystyle\sum\limits_{\sigma\in A}}
\sigma\widehat{x}_{\sigma}%
\]
that is analogous to an element in the Lie algebra of the translation group.
\ $X$ resides in an algebra that is formed by taking the direct product of
$A(F_{N}),$where $\sigma\in A(F_{N}),$ with the group algebra of T$_{x}(A)$
over the complex numbers, where $\widehat{x}_{\sigma}\in$ group algebra of
T$_{x}(A)$. \ Properties of this direct product algebra are developed only as
needed in the following; one may initially view\ the expression for $X$, and
similar expressions below,\ as a formal expression.

Consider the action of $t_{x}^{h}$ on a function $f=%
{\displaystyle\sum\limits_{k\in A}}
f(k)t_{x}^{k}$ in the group algebra of \ T$_{x}(A)$ over the complex numbers
\[
t_{x}^{h}f=\exp(\frac{2\pi i}{N}\text{Tr}(h%
{\displaystyle\sum\limits_{\sigma\in A}}
\sigma\widehat{x}_{\sigma}))%
{\displaystyle\sum\limits_{k\in A}}
f(k)t_{x}^{k}%
\]
Expand out the exponential and let $t_{x}^{k}$ act on $\widehat{x}_{\sigma}$
to obtain%
\[
t_{x}^{h}f=%
{\displaystyle\sum\limits_{\sigma\in A}}
{\displaystyle\sum\limits_{m=0}^{\infty}}
\frac{1}{m!}\left(  \frac{2\pi i}{N}\text{Tr}h\sigma\right)  ^{m}%
{\displaystyle\sum\limits_{k\in A}}
\exp(\frac{2\pi i}{N}\text{Tr}\sigma k)f(k)\widehat{x}_{\sigma}%
\]
Now expanding $\widehat{x}_{\sigma}$ using Eq. $\ref{estate for tx}$ we obtain
\
\[
t_{x}^{h}f=%
{\displaystyle\sum\limits_{l\in A}}
{\displaystyle\sum\limits_{m=0}^{\infty}}
\frac{1}{m!}\left(  \frac{1}{V}%
{\displaystyle\sum\limits_{\sigma,k\in A}}
\left(  \frac{2\pi i}{N}\text{Tr}h\sigma\right)  ^{m}\exp(\frac{2\pi i}%
{N}\text{Tr}\sigma(k-l))f(k)\right)  t_{x}^{l}%
\]
Since we also have $t_{x}^{h}f=%
{\displaystyle\sum\limits_{l\in A}}
f(l-h)t_{x}^{l}$ we conclude%
\begin{align}
f(l-h)  &  =%
{\displaystyle\sum\limits_{m=0}^{\infty}}
\frac{1}{m!}\left(  \frac{2\pi i}{N}\right)  ^{m}\left(  \frac{1}{V}%
{\displaystyle\sum\limits_{\sigma,k\in A}}
\left(  \text{Tr}h\sigma\right)  ^{m}\exp(\frac{2\pi i}{N}\text{Tr}%
\sigma(k-l))f(k)\right) \nonumber\\
&  \label{TS}%
\end{align}
This is our analog to the Taylor series expansion of a function about a
position $l$ by a displacement $h$. \ We use the first order term in this
expression to define an operator that is analogous to the directional
derivative or gradient operator
\begin{equation}
\operatorname{grad}f(l)=^{\text{def}}\frac{1}{V}%
{\displaystyle\sum\limits_{\sigma,k\in A}}
\sigma\exp(\frac{2\pi i}{N}\text{Tr}\sigma(k-l))f(k) \label{graddef}%
\end{equation}
This is our counterpart to Eq. 2.31 in Folland (\cite{F1} p 93) that describes
the Kohn-Nirenberg correspondence in the theory of pseudo-differential
operators. \ $\operatorname{grad}f(l)$ resides in the algebra formed by the
direct product of $A(F_{N}),$where $\sigma\in A(F_{N}),$ with the complex
numbers. \ The $i^{th}$component of $\operatorname{grad}f(l)$ provides our
definition of the partial derivative operation%
\begin{align*}
\frac{\partial}{\partial l_{i}}f(l)  &  =^{\text{def}}\operatorname{grad}%
f(l)\mid_{i}\\
&  =\frac{1}{V}%
{\displaystyle\sum\limits_{\sigma,k\in A}}
\sigma_{i}\exp(\frac{2\pi i}{N}\text{Tr}\sigma(k-l))f(k)
\end{align*}
Using these definitions, we have
\[
\operatorname{grad}f(l)=%
{\displaystyle\sum\limits_{e_{i}}}
e_{i}\frac{\partial}{\partial l_{i}}f(l)
\]
\ One can develop the algebra of $\operatorname{grad}$ in a manner similar to
the treatment of Doran and Lasenby (\cite{D}, p 168) in the context of
geometric algebra. \ We limit the development in the following to our
immediate applications.

Iterating the expression for Tr$\left(  h\operatorname{grad}\right)  f(l)$we
find%
\[
\left(  \text{Tr}\left(  h\operatorname{grad}\right)  \right)  ^{m}%
f(l)=\frac{1}{V}%
{\displaystyle\sum\limits_{\sigma,k\in A}}
(\text{Tr}h\sigma)^{m}\exp(\frac{2\pi i}{N}\text{Tr}(\sigma(k-l)))f(k)
\]
Exponentiating Tr$\left(  h\operatorname{grad}\right)  $ we obtain
\begin{equation}
f(l-h)=\exp(\frac{2\pi i}{N}\text{Tr}(h\operatorname{grad}))f(l) \label{grad}%
\end{equation}
We derive rules for differentiation by applying Eq. \ref{grad}$\ $to a chosen
function and equating terms according to their powers of $h$ on the two sides
of the equation. \ E.g., applying $\exp(\frac{2\pi i}{N}$%
Tr$(h\operatorname{grad}))$ to the function $f(l)=f_{1}(l)f_{2}(l)$
\begin{align*}
\exp(\frac{2\pi i}{N}\text{Tr}(h\operatorname{grad}))f(l)  &  =f_{1}%
(l-h)f_{2}(l-h)\\
&  =[\exp(\frac{2\pi i}{N}\text{Tr}(h\operatorname{grad}))f_{1}(l)][\exp
(\frac{2\pi i}{N}\text{Tr}(h\operatorname{grad}))f_{2}(l)]
\end{align*}
and equating the first order terms in $h$ we obtain the Leibnitz rule.

Comparing Eq. \ref{grad} with Eq. \ref{action of tx} we have the exponential
form for the operater $t_{x}^{h}$
\begin{equation}
t_{x}^{h}f=%
{\displaystyle\sum\limits_{k\in A}}
\exp(\frac{2\pi i}{N}\text{Tr}h\operatorname{grad})f(k)t_{x}^{k}I
\label{txdif}%
\end{equation}

We calculate the conjugate reversal of grad and iterate with grad to obtain%
\begin{align}
\operatorname{grad}\operatorname{grad}^{\ast}f(l)  &  =\frac{1}{V}%
{\displaystyle\sum\limits_{\sigma,k\in A}}
\sigma\sigma^{\ast}\exp(\frac{2\pi i}{N}\text{Tr}\sigma(k-l))f(k)\\
&  \label{grgr}%
\end{align}
Just as for $\operatorname{grad}$ in Eq. \ref{grad} we can form the
exponential of the trace of this operator. \ Comparing the result with
expression Eq. \ref{td action} we find
\begin{equation}
t_{\ d}^{c}f=%
{\displaystyle\sum\limits_{l\in A}}
\exp(\frac{-2\pi i}{N}\frac{c}{4\omega_{0}}\text{Tr}\operatorname{grad}%
\operatorname{grad}^{\ast})f(l)t_{x}^{l}I \label{td d action}%
\end{equation}

\subsection{Eigenstates of group actions}

We use these derivative operations to reexpress eigenstates of group actions
in the Schr\"{o}dinger-Weil representation in a differential form.

Projecting with an eigenstate of the translation operator
\begin{align*}
\widehat{x}_{\sigma}f  &  =\frac{1}{V}%
{\displaystyle\sum\limits_{h\in S}}
\exp(\frac{-2\pi i}{N}\text{Tr}h\sigma)t_{x}^{h}f\\
&  =\frac{1}{V}%
{\displaystyle\sum\limits_{h,k\in S}}
\exp(\frac{-2\pi i}{N}\text{Tr}h(\sigma-\operatorname{grad}))f(k)t_{x}^{k}I\\
&  =%
{\displaystyle\sum\limits_{k\in S}}
\delta(\operatorname{grad}-\sigma)f(k)t_{x}^{k}I
\end{align*}
Therefore eigenstates $\widehat{x}_{\sigma}f$ satisfy the operator identity
\[
0=(\operatorname{grad}-\sigma)f(k).
\]

Projecting with $\widehat{d}_{\eta}=\frac{1}{N}%
{\displaystyle\sum\limits_{h\in F_{N}}}
\exp(\frac{2\pi i}{N}h\eta)t_{d}^{h}$ and using Eq. \ref{td d action} for the
derivative form for $t_{d}^{h}$ we find that eigenstates $\widehat{d}_{\eta}f$
\ satisfy
\begin{equation}
0=(\text{Tr}(\operatorname{grad}\operatorname{grad}^{\ast})+4\omega_{0}%
\eta)f(k) \label{KG}%
\end{equation}
This is the Klein-Gordan equation for the configuation space $A(F_{N})$. \ We
can verify this expression by writing $f$ in terms of its fourier components%
\[
f=%
{\displaystyle\sum\limits_{\sigma\in S}}
\widetilde{f}(\sigma)\widehat{x}_{\sigma}I
\]
for $\widetilde{f}(\sigma)=%
{\displaystyle\sum\limits_{k\in S}}
\exp(\frac{2\pi i}{N}$Tr$k\sigma)f(k).$ \ Since
\[
\widehat{d}_{\eta}\widehat{x}_{\sigma}I=\delta(\eta+\frac{1}{4\omega_{0}%
}\text{Tr}(\sigma\sigma^{\ast}))\widehat{x}_{\sigma}I
\]
we have%
\begin{align*}
\widehat{d}_{\eta}f  &  =%
{\displaystyle\sum\limits_{\sigma\in S;Tr(\sigma\sigma^{\ast})=-4\omega
_{0}\eta}}
\widetilde{f}(\sigma)\widehat{x}_{\sigma}I\\
&  =\frac{1}{V}%
{\displaystyle\sum\limits_{k\in S}}
(%
{\displaystyle\sum\limits_{\sigma\in S;\text{Tr}(\sigma\sigma^{\ast}%
)=-4\omega_{0}\eta}}
\widetilde{f}(\sigma)\exp(\frac{-2\pi i}{N}\text{Tr}k\sigma))t_{x}^{k}I
\end{align*}
Applying $\operatorname{grad}\operatorname{grad}^{\ast}$ to the eigenfunction
$%
{\displaystyle\sum\limits_{\sigma\in S\text{;Tr}(\sigma\sigma^{\ast}%
)=-4\omega_{0}\eta}}
\widetilde{f}(\sigma)\exp(\frac{-2\pi i}{N}$Tr$k\sigma)$ returns Eq. \ref{KG}.

\ The operator $t_{u}^{h}\in$SL$(2,F_{N})$ that is conjugate to $t_{d}^{-h}$
\[
t_{u}^{h}=jt_{d}^{-h}j^{-1}%
\]
\ \ \ \ \ \ \ \ \ \ \ \ \ \ \ \ \ \ \ \ \ \ \ \ \ \ \ \ \ \ \ \ \ \ \ \ \ \ \ \ \ \ \ \ \ \ \ \ \ \ \ \ \ \ \ \ \ \ \ \ \ \ \ \ \ \ \ \ \ \ \ \ \ \ \ \ \ \ \ \ \ \ \ \ \ \ \ \ \ \ \ \ \ \ \ \ \ \ \ \ \ \ \ \ \ \ \ \ \ \ \ \ \ \ \ \ \ \ \ \ \ \ \ \ \ \ \ \ \ \ \ \ \ \ \ \ \ \ \ \ \ \ \ \ \ \ \ \ \ \ \ \ \ \ \ \ \ \ \ \ \ \ \ \ \ \ \ \ \ \ \ \ \ \ \ \ \ \ \ \ \ \ \ \ \ \ \ \ \ \ \ \ \ \ \ \ \ \ \ \ \ \ \ \ \ \ \ \ \ \ \ \ \ \ \ \ \ \ \ \ \ \ \ \ \ \ \ \ \ \ \ \ \ \ \ \ \ \ \ \ \ \ \ \ \ \ \ \ \ \ \ \ \ \ \ \ \ \ \ \ \ \ \ \ \ \ \ \ \ \ \ \ \ \ \ \ \ \ \ \ \ \ \ \ \ \ \ \ \ \ \ \ \ \ \ \ \ \ \ \ \ \ \ \ \ \ \ \ \ \ \ \ \ \ \ \ \ \ \ \ \ \ \ \ \ \ \ \ \ \ \ \ \ \ \ \ \ \ \ \ \ \ \ \ \ \ \ \ \ \ \ \ \ \ \ \ \ \ \ \ \ \ \ \ \ \ \ \ \ \ \ \ \ \ \ \ \ \ \ \ \ \ \ \ \ \ \ \ \ \ \ \ \ \ \ \ \ \ \ \ \ \ \ \ \ \ \ \ \ \ \ \ \ \ \ \ \ \ \ \ \ \ \ \ \ \ \ \ \ \ \ \ \ \ \ \ \ \ \ \ \ \ \ \ \ \ \ \ \ \ \ \ \ \ \ \ \ \ \ \ \ \ \ \ \ \ \ \ \ \ \ \ \ \ \ \ \ \ \ \ \ \ \ \ \ \ \ \ \ \ \ \ \ \ \ \ \ \ \ \ \ \ \ \ \ \ \ \ \ \ \ \ \ \ \ \ \ \ \ \ \ \ \ \ \ \ \ \ \ \ \ \ \ \ \ \ \ \ \ \ \ \ \ \ \ \ \ \ \ \ \ \ \ \ \ \ \ \ \ \ \ \ \ \ \ \ \ \ \ \ \ \ \ \ \ \ \ \ \ \ \ \ \ \ \ \ \ \ \ \ \ \ \ \ \ \ \ \ \ \ \ \ \ \ \ \ \ \ \ \ \ \ \ \ \ \ \ \ \ \ \ \ \ \ \ \ \ \ \ \ \ \ \ \ \ \ \ \ \ \ \ \ \ \ \ \ \ \ \ \ \ \ \ \ \ \ \ \ \ \ \ \ \ \ \ \ \ \ \ \ \ \ \ \ \ \ \ \ \ \ \ \ \ \ \ \ \ \ \ \ \ \ \ \ \ \ \ \ \ \ \ \ \ \ \ \ \ \ \ \ \ \ \ \ \ \ \ \ \ \ \ \ \ \ \ \ \ \ \ \ \ \ \ \ \ \ \ \ \ \ \ \ \ \ \ \ \ \ \ \ \ \ \ \ \ \ \ \ \ \ \ \ \ \ \ \ \ \ \ \ \ \ \ \ \ \ \ \ \ \ \ \ \ \ \ \ \ \ \ \ \ \ \ \ \ \ \ \ \ \ \ \ \ \ \ \ \ \ \ \ \ \ \ \ \ \ \ \ \ \ \ \ \ \ \ \ \ \ \ \ \ \ \ \ \ \ \ \ \ \ \ \ \ \ \ \ \ \ \ \ \ \ \ \ \ \ \ \ \ \ \ \ \ \ \ \ \ \ \ \ \ \ \ \ \ \ \ \ \ \ \ \ \ \ \ \ \ \ \ \ \ \ \ \ \ \ \ \ \ \ \ \ \ \ \ \ \ \ \ \ \ \ \ \ \ \ \ \ \ \ \ \ \ \ \ \ \ \ \ \ \ \ \ \ \ \ \ \ \ \ \ \ \ \ \ \ \ \ \ \ \ \ \ \ \ \ \ \ \ \ \ \ \ \ \ \ \ \ \ \ \ \ \ \ \ \ \ \ \ \ \ \ \ \ \ \ \ \ \ \ \ \ \ \ \ \ \ \ \ \ \ \ \ \ \ \ \ \ \ \ \ \ \ \ \ \ \ \ \ \ \ \ \ \ \ \ \ \ \ \ \ \ \ \ \ \ \ \ \ \ \ \ \ \ \ \ \ \ \ \ \ \ \ \ \ \ \ \ \ \ \ \ \ \ \ \ \ \ \ \ \ \ \ \ \ \ \ \ \ \ \ \ \ \ \ \ \ \ \ \ \ \ \ \ \ \ \ \ \ \ \ \ \ \ \ \ \ \ \ \ \ \ \ \ \ \ \ \ \ \ \ \ \ \ \ \ \ \ \ \ \ \ \ \ \ \ \ \ \ \ \ \ \ \ \ \ \ \ \ \ \ \ \ \ \ \ \ \ \ \ \ \ \ \ \ \ \ \ \ \ \ \ \ \ \ \ \ \ \ \ \ \ \ \ \ \ \ \ \ \ \ \ \ \ \ \ \ \ \ \ \ \ \ \ \ \ \ \ \ \ \ \ \ \ \ \ \ \ \ \ \ \ \ \ \ \ \ \ \ \ \ \ \ \ \ \ \ \ \ \ \ \ \ \ \ \ \ \ \ \ \ \ \ \ \ \ \ \ \ \ \ \ \ \ \ \ \ \ \ \ \ \ \ \ \ \ \ \ \ \ \ \ \ \ \ \ \ \ \ \ \ \ \ \ \ \ \ \ \ \ \ \ \ \ \ \ \ \ \ \ \ \ \ \ \ \ \ \ \ \ \ \ \ \ \ \ \ \ \ \ \ \ \ \ \ \ \ \ \ \ \ \ \ \ \ \ \ \ \ \ \ \ \ \ \ \ \ \ \ \ \ \ \ \ \ \ \ \ \ \ \ \ \ \ \ \ \ \ \ \ \ \ \ \ \ \ \ \ \ \ \ \ \ \ \ \ \ \ \ \ \ \ \ \ \ \ \ \ \ \ \ \ \ \ \ \ \ \ \ \ \ \ \ \ \ \ \ \ \ \ \ \ \ \ \ \ \ \ \ \ \ \ \ \ \ \ \ \ \ \ \ \ \ \ \ \ \ \ \ \ \ \ \ \ \ \ \ \ \ \ \ \ \ \ \ \ \ \ \ \ \ \ \ \ \ \ \ \ \ \ \ \ \ \ \ \ \ \ \ \ \ \ \ \ \ \ \ \ \ \ \ \ \ \ \ \ \ \ \ \ \ \ \ \ \ \ \ \ \ \ \ \ \ \ \ \ \ \ \ \ \ \ \ \ \ \ \ \ \ \ \ \ \ \ \ \ \ \ \ \ \ \ \ \ \ \ \ \ \ \ \ \ \ \ \ \ \ \ \ \ \ \ \ \ \ \ \ \ \ \ \ \ \ \ \ \ \ \ \ \ \ \ \ \ \ \ \ \ \ \ \ \ \ \ \ \ \ \ \ \ \ \ \ \ \ \ \ \ \ \ \ \ \ \ \ \ \ \ \ \ \ \ \ \ \ \ \ \ \ \ \ \ \ \ \ \ \ \ \ \ \ \ \ \ \ \ \ \ \ \ \ \ \ \ \ \ \ \ \ \ \ \ \ \ \ \ \ \ \ \ \ \ \ \ \ \ \ \ \ \ \ \ \ \ \ \ \ \ \ \ \ \ \ \ \ \ \ \ \ \ \ \ \ \ \ \ \ \ \ \ \ \ \ \ \ \ \ \ \ \ \ \ \ \ \ \ \ \ \ \ \ \ \ \ \ \ \ \ \ \ \ \ \ \ \ \ \ \ \ \ \ \ \ \ \ \ \ \ \ \ \ \ \ \ \ \ \ \ \ \ \ \ \ \ \ \ \ \ \ \ \ \ \ \ \ \ \ \ \ \ \ \ \ \ \ \ \ \ \ \ \ \ \ \ \ \ \ \ \ \ \ \ \ \ \ \ \ \ \ \ \ \ \ \ \ \ \ \ \ \ \ \ \ \ \ \ \ \ \ \ \ \ \ \ \ \ \ \ \ \ \ \ \ \ \ \ \ \ \ \ \ \ \ \ \ \ \ \ \ \ \ \ \ \ \ \ \ \ \ \ \ \ \ \ \ \ \ \ \ \ \ \ \ \ \ \ \ \ \ \ \ \ \ \ \ \ \ \ \ \ \ \ \ \ \ \ \ \ \ \ \ \ \ \ \ \ \ \ \ \ \ \ \ \ \ \ \ \ \ \ \ \ \ \ \ \ \ \ \ \ \ \ \ \ \ \ \ \ \ \ \ \ \ \ \ \ \ \ \ \ \ \ \ \ \ \ \ \ \ \ \ \ \ \ \ \ \ \ \ \ \ \ \ \ \ \ \ \ \ \ \ \ \ \ \ \ \ \ \ \ \ \ \ \ \ \ \ \ \ \ \ \ \ \ \ \ \ \ \ \ \ \ \ \ \ \ \ \ \ \ \ \ \ \ \ \ \ \ \ \ \ \ \ \ \ \ \ \ \ \ \ \ \ \ \ \ \ \ \ \ \ \ \ \ \ \ \ \ \ \ \ \ \ \ \ \ \ \ \ \ \ \ \ \ \ \ \ \ \ \ \ \ \ \ \ \ \ \ \ \ \ \ \ \ \ \ \ \ \ \ \ \ \ \ \ \ \ \ \ \ \ \ \ \ \ \ \ \ \ \ \ \ \ \ \ \ \ \ \ \ \ \ \ \ \ \ \ \ \ \ \ \ \ \ \ \ \ \ \ \ \ \ \ \ \ \ \ \ \ \ \ \ \ \ \ \ \ \ \ \ \ \ \ \ \ \ \ \ \ \ \ \ \ \ \ \ \ \ \ \ \ \ \ \ \ \ \ \ \ \ \ \ \ \ \ \ \ \ \ \ \ \ \ \ \ \ \ \ \ \ \ \ \ \ \ \ \ \ \ \ \ \ \ \ \ \ \ \ \ \ \ \ \ \ \ \ \ \ \ \ \ \ \ \ \ \ \ \ \ \ \ \ \ \ \ \ \ \ \ \ \ \ \ \ \ \ \ \ \ \ \ \ \ \ \ \ \ \ \ \ \ \ \ \ \ \ \ \ \ \ \ \ \ \ \ \ \ \ \ \ \ \ \ \ \ \ \ \ \ \ \ \ \ \ \ \ \ \ \ \ \ \ \ \ \ \ \ \ \ \ \ \ \ \ \ \ \ \ \ \ \ \ \ \ \ \ \ \ \ \ \ \ \ \ \ \ \ \ \ \ \ \ \ \ \ \ \ \ \ \ \ \ \ \ \ \ \ \ \ \ \ \ \ \ \ \ \ \ \ \ \ \ \ \ \ \ \ \ \ \ \ \ \ \ \ \ \ \ \ \ \ \ \ \ \ \ \ \ \ \ \ \ \ \ \ \ \ \ \ \ \ \ \ \ \ \ \ \ \ \ \ \ \ \ \ \ \ \ \ \ \ \ \ \ \ \ \ \ \ \ \ \ \ \ \ \ \ \ \ \ \ \ \ \ \ \ \ \ \ \ \ \ \ \ \ \ \ \ \ \ \ \ \ \ \ \ \ \ \ \ \ \ \ \ \ \ \ \ \ \ \ \ \ \ \ \ \ \ \ \ \ \ \ \ \ \ \ \ \ \ \ \ \ \ \ \ \ \ \ \ \ \ \ \ \ \ \ \ \ \ \ \ \ \ \ \ \ \ \ \ \ \ \ \ \ \ \ \ \ \ \ \ \ \ \ \ \ \ \ \ \ \ \ \ \ \ \ \ \ \ \ \ \ \ \ \ \ \ \ \ \ \ \ \ \ \ \ \ \ \ \ \ \ \ \ \ \ \ \ \ \ \ \ \ \ \ \ \ \ \ \ \ \ \ \ \ \ \ \ \ \ \ \ \ \ \ \ \ \ \ \ \ \ \ \ \ \ \ \ \ \ \ \ \ \ \ \ \ \ \ \ \ \ \ \ \ \ \ \ \ \ \ \ \ \ \ \ \ \ \ \ \ \ \ \ \ \ \ \ \ \ \ \ \ \ \ \ \ \ \ \ \ \ \ \ \ \ \ \ \ \ \ \ \ \ \ \ \ \ \ \ \ \ \ \ \ \ \ \ \ \ \ \ \ \ \ \ \ \ \ \ \ \ \ \ \ \ \ \ \ \ \ \ \ \ \ \ \ \ \ \ \ \ \ \ \ \ \ \ \ \ \ \ \ \ \ \ \ \ \ \ \ \ \ \ \ \ \ \ \ \ \ \ \ \ \ \ \ \ \ \ \ \ \ \ \ \ \ \ \ \ \ \ \ \ \ \ \ \ \ \ \ \ \ \ \ \ \ \ \ \ \ \ \ \ \ \ \ \ \ \ \ \ \ \ \ \ \ \ \ \ \ \ \ \ \ \ \ \ \ \ \ \ \ \ \ \ \ \ \ \ \ \ \ \ \ \ \ \ \ \ \ \ \ \ \ \ \ \ \ \ \ \ \ \ \ \ \ \ \ \ \ \ \ \ \ \ \ \ \ \ \ \ \ \ \ \ \ \ \ \ \ \ \ \ \ \ \ \ \ \ \ \ \ \ \ \ \ \ \ \ \ \ \ \ \ \ \ \ \ \ \ \ \ \ \ \ \ \ \ \ \ \ \ \ \ \ \ \ \ \ \ \ \ \ \ \ \ \ \ \ \ \ \ \ \ \ \ \ \ \ \ \ \ \ \ \ \ \ \ \ \ \ \ \ \ \ \ \ \ \ \ \ \ \ \ \ \ \ \ \ \ \ \ \ \ \ \ \ \ \ \ \ \ \ \ \ \ \ \ \ \ \ \ \ \ \ \ \ \ \ \ \ \ \ \ \ \ \ \ \ \ \ \ \ \ \ \ \ \ \ \ \ \ \ \ \ \ \ \ \ \ \ \ \ \ \ \ \ \ \ \ \ \ \ \ \ \ \ \ \ \ \ \ \ \ \ \ \ \ \ \ \ \ \ \ \ \ \ \ \ \ \ \ \ \ \ \ \ \ \ \ \ \ \ \ \ \ \ \ \ \ \ \ \ \ \ \ \ \ \ \ \ \ \ \ \ \ \ \ \ \ \ \ \ \ \ \ \ \ \ \ \ \ \ \ \ \ \ \ \ \ \ \ \ \ \ \ \ \ \ \ \ \ \ \ \ \ \ \ \ \ \ \ \ \ \ \ \ \ \ \ \ \ \ \ \ \ \ \ \ \ \ \ \ \ \ \ \ \ \ \ \ \ \ \ \ \ \ \ \ \ \ \ \ \ \ \ \ \ \ \ \ \ \ \ \ \ \ \ \ \ \ \ \ \ \ \ \ \ \ \ \ \ \ \ \ \ \ \ \ \ \ \ \ \ \ \ \ \ \ \ \ \ \ \ \ \ \ \ \ \ \ \ \ \ \ \ \ \ \ \ \ \ \ \ \ \ \ \ \ \ \ \ \ \ \ \ \ \ \ \ \ \ \ \ \ \ \ \ \ \ \ \ \ \ \ \ \ \ \ \ \ \ \ \ \ \ \ \ \ \ \ \ \ \ \ \ \ \ \ \ \ \ \ \ \ \ \ \ \ \ \ \ \ \ \ \ \ \ \ \ \ \ \ \ \ \ \ \ \ \ \ \ \ \ \ \ \ \ \ \ \ \ \ \ \ \ \ \ \ \ \ \ \ \ \ \ \ \ \ \ \ \ \ \ \ \ \ \ \ \ \ \ \ \ \ \ \ \ \ \ \ \ \ \ \ \ \ \ \ \ \ \ \ \ \ \ \ \ \ \ \ \ \ \ \ \ \ \ \ \ \ \ \ \ \ \ \ \ \ \ \ \ \ \ \ \ \ \ \ \ \ \ \ \ \ \ \ \ \ \ \ \ \ \ \ \ \ \ \ \ \ \ \ \ \ \ \ \ \ \ \ \ \ \ \ \ \ \ \ \ \ \ \ \ \ \ \ \ \ \ \ \ \ \ \ \ \ \ \ \ \ \ \ \ \ \ \ \ \ \ \ \ \ \ \ \ \ \ \ \ \ \ \ \ \ \ \ \ \ \ \ \ \ \ \ \ \ \ \ \ \ \ \ \ \ \ \ \ \ \ \ \ \ \ \ \ \ \ \ \ \ \ \ \ \ \ \ \ \ \ \ \ \ \ \ \ \ \ \ \ \ \ \ \ \ \ \ \ \ \ \ \ \ \ \ \ \ \ \ \ \ \ \ \ \ \ \ \ \ \ \ \ \ \ \ \ \ \ \ \ \ \ \ \ \ \ \ \ \ \ \ \ \ \ \ \ \ \ \ \ \ \ \ \ \ \ \ \ \ \ \ \ \ \ \ \ \ \ \ \ \ \ \ \ \ \ \ \ \ \ \ \ \ \ \ \ \ \ \ \ \ \ \ \ \ \ \ \ \ \ \ \ \ \ \ \ \ \ \ \ \ \ \ \ \ \ \ \ \ \ \ \ \ \ \ \ \ \ \ \ \ \ \ \ \ \ \ \ \ \ \ \ \ \ \ \ \ \ \ \ \ \ \ \ \ \ \ \ \ \ \ \ \ \ \ \ \ \ \ \ \ \ \ \ \ \ \ \ \ \ \ \ \ \ \ \ \ \ \ \ \ \ \ \ \ \ \ \ \ \ \ \ \ \ \ \ \ \ \ \ \ \ \ \ \ \ \ \ \ \ \ \ \ \ \ \ \ \ \ \ \ \ \ \ \ \ \ \ \ \ \ \ \ \ \ \ \ \ \ \ \ \ \ \ \ \ \ \ \ \ \ \ \ \ \ \ \ \ \ \ \ \ \ \ \ \ \ \ \ \ \ \ \ \ \ \ \ \ \ \ \ \ \ \ \ \ \ \ \ \ \ \ \ \ \ \ \ \ \ \ \ \ \ \ \ \ \ \ \ \ \ \ \ \ \ \ \ \ \ \ \ \ \ \ \ \ \ \ \ \ \ \ \ \ \ \ \ \ \ \ \ \ \ \ \ \ \ \ \ \ \ \ \ \ \ \ \ \ \ \ \ \ \ \ \ \ \ \ \ \ \ \ \ \ \ \ \ \ \ \ \ \ \ \ \ \ \ \ \ \ \ \ \ \ \ \ \ \ \ \ \ \ \ \ \ \ \ \ \ \ \ \ \ \ \ \ \ \ \ \ \ \ \ \ \ \ \ \ \ \ \ \ \ \ \ \ \ \ \ \ \ \ \ \ \ \ \ \ \ \ \ \ \ \ \ \ \ \ \ \ \ \ \ \ \ \ \ \ \ \ \ \ \ \ \ \ \ \ \ \ \ \ \ \ \ \ \ \ \ \ \ \ \ \ \ \ \ \ \ \ \ \ \ \ \ \ \ \ \ \ \ \ \ \ \ \ \ \ \ \ \ \ \ \ \ \ \ \ \ \ \ \ \ \ \ \ \ \ \ \ \ \ \ \ \ \ \ \ \ \ \ \ \ \ \ \ \ \ \ \ \ \ \ \ \ \ \ \ \ \ \ \ \ \ \ \ \ \ \ \ \ \ \ \ \ \ \ \ \ \ \ \ \ \ \ \ \ \ \ \ \ \ \ \ \ \ \ \ \ \ \ \ \ \ \ \ \ \ \ \ \ \ \ \ \ \ \ \ \ \ \ \ \ \ \ \ \ \ \ \ \ \ \ \ \ \ \ \ \ \ \ \ \ \ \ \ \ \ \ \ \ \ \ \ \ \ \ \ \ \ \ \ \ \ \ \ \ \ \ \ \ \ \ \ \ \ \ \ \ \ \ \ \ \ \ \ \ \ \ \ \ \ \ \ \ \ \ \ \ \ \ \ \ \ \ \ \ \ \ \ \ \ \ \ \ \ \ \ \ \ \ \ \ \ \ \ \ \ \ \ \ \ \ \ \ \ \ \ \ \ \ \ \ \ \ \ \ \ \ \ \ \ \ \ \ \ \ \ \ \ \ \ \ \ \ \ \ \ \ \ \ \ \ \ \ \ \ \ \ \ \ \ \ \ \ \ \ \ \ \ \ \ \ \ \ \ \ \ \ \ \ \ \ \ \ \ \ \ \ \ \ \ \ \ \ \ \ \ \ \ \ \ \ \ \ \ \ \ \ \ \ \ \ \ \ \ \ \ \ \ \ \ \ \ \ \ \ \ \ \ \ \ \ \ \ \ \ \ \ \ \ \ \ \ \ \ \ \ \ \ \ \ \ \ \ \ \ \ \ \ \ \ \ \ \ \ \ \ \ \ \ \ \ \ \ \ \ \ \ \ \ \ \ \ \ \ \ \ \ \ \ \ \ \ \ \ \ \ \ \ \ \ \ \ \ \ \ \ \ \ \ \ \ \ \ \ \ \ \ \ \ \ \ \ \ \ \ \ \ \ \ \ \ \ \ \ \ \ \ \ \ \ \ \ \ \ \ \ \ \ \ \ \ \ \ \ \ \ \ \ \ \ \ \ \ \ \ \ \ \ \ \ \ \ \ \ \ \ \ \ \ \ \ \ \ \ \ \ \ \ \ \ \ \ \ \ \ \ \ \ \ \ \ \ \ \ \ \ \ \ \ \ \ \ \ \ \ \ \ \ \ \ \ \ \ \ \ \ \ \ \ \ \ \ \ \ \ \ \ \ \ \ \ \ \ \ \ \ \ \ \ \ \ \ \ \ \ \ \ \ \ \ \ \ \ \ \ \ \ \ \ \ \ \ \ \ \ \ \ \ \ \ \ \ \ \ \ \ \ \ \ \ \ \ \ \ \ \ \ \ \ \ \ \ \ \ \ \ \ \ \ \ \ \ \ \ \ \ \ \ \ \ \ \ \ \ \ \ \ \ \ \ \ \ \ \ \ \ \ \ \ \ \ \ \ \ \ \ \ \ \ \ \ \ \ \ \ \ \ \ \ \ \ \ \ \ \ \ \ \ \ \ \ \ \ \ \ \ \ \ \ \ \ \ \ \ \ \ \ \ \ \ \ \ \ \ \ \ \ \ \ \ \ \ \ \ \ \ \ \ \ \ \ \ \ \ \ \ \ \ \ \ \ \ \ \ \ \ \ \ \ \ \ \ \ \ \ \ \ \ \ \ \ \ \ \ \ \ \ \ \ \ \ \ \ \ \ \ \ \ \ \ \ \ \ \ \ \ \ \ \ \ \ \ \ \ \ \ \ \ \ \ \ \ \ \ \ \ \ \ \ \ \ \ \ \ \ \ \ \ \ \ \ \ \ \ \ \ \ \ \ \ \ \ \ \ \ \ \ \ \ \ \ \ \ \ \ \ \ \ \ \ \ \ \ \ \ \ \ \ \ \ \ \ \ \ \ \ \ \ \ \ \ \ \ \ \ \ \ \ \ \ \ \ \ \ \ \ \ \ \ \ \ \ \ \ \ \ \ \ \ \ \ \ \ \ \ \ \ \ \ \ \ \ \ \ \ \ \ \ \ \ \ \ \ \ \ \ \ \ \ \ \ \ \ \ \ \ \ \ \ \ \ \ \ \ \ \ \ \ \ \ \ \ \ \ \ \ \ \ \ \ \ \ \ \ \ \ \ \ \ \ \ \ \ \ \ \ \ \ \ \ \ \ \ \ \ \ \ \ \ \ \ \ \ \ \ \ \ \ \ \ \ \ \ \ \ \ \ \ \ \ \ \ \ \ \ \ \ \ \ \ \ \ \ \ \ \ \ \ \ \ \ \ \ \ \ \ \ \ \ \ \ \ \ \ \ \ \ \ \ \ \ \ \ \ \ \ \ \ \ \ \ \ \ \ \ \ \ \ \ \ \ \ \ \ \ \ \ \ \ \ \ \ \ \ \ \ \ \ \ \ \ \ \ \ \ \ \ \ \ \ \ \ \ \ \ \ \ \ \ \ \ \ \ \ \ \ \ \ \ \ \ \ \ \ \ \ \ \ \ \ \ \ \ \ \ \ \ \ \ \ \ \ \ \ \ \ \ \ \ \ \ \ \ \ \ \ \ \ \ \ \ \ \ \ \ \ \ \ \ \ \ \ \ \ \ \ \ \ \ \ \ \ \ \ \ \ \ \ \ \ \ \ \ \ \ \ \ \ \ \ \ \ \ \ \ \ \ \ \ \ \ \ \ \ \ \ \ \ \ \ \ \ \ \ \ \ \ \ \ \ \ \ \ \ \ \ \ \ \ \ \ \ \ \ \ \ \ \ \ \ \ \ \ \ \ \ \ \ \ \ \ \ \ \ \ \ \ \ \ \ \ \ \ \ \ \ \ \ \ \ \ \ \ \ \ \ \ \ \ \ \ \ \ \ \ \ \ \ \ \ \ \ \ \ \ \ \ \ \ \ \ \ \ \ \ \ \ \ \ \ \ \ \ \ \ \ \ \ \ \ \ \ \ \ \ \ \ \ \ \ \ \ \ \ \ \ \ \ \ \ \ \ \ \ \ \ \ \ \ \ \ \ \ \ \ \ \ \ \ \ \ \ \ \ \ \ \ \ \ \ \ \ \ \ \ \ \ \ \ \ \ \ \ \ \ \ \ \ \ \ \ \ \ \ \ \ \ \ \ \ \ \ \ \ \ \ \ \ \ \ \ \ \ \ \ \ \ \ \ \ \ \ \ \ \ \ \ \ \ \ \ \ \ \ \ \ \ \ \ \ \ \ \ \ \ \ \ \ \ \ \ \ \ \ \ \ \ \ \ \ \ \ \ \ \ \ \ \ \ \ \ \ \ \ \ \ \ \ \ \ \ \ \ \ \ \ \ \ \ \ \ \ \ \ \ \ \ \ \ \ \ \ \ \ \ \ \ \ \ \ \ \ \ \ \ \ \ \ \ \ \ \ \ \ \ \ \ \ \ \ \ \ \ \ \ \ \ \ \ \ \ \ \ \ \ \ \ \ \ \ \ \ \ \ \ \ \ \ \ \ \ \ \ \ \ \ \ \ \ \ \ \ \ \ \ \ \ \ \ \ \ \ \ \ \ \ \ \ \ \ \ \ \ \ \ \ \ \ \ \ \ \ \ \ \ \ \ \ \ \ \ \ \ \ \ \ \ \ \ \ \ \ \ \ \ \ \ \ \ \ \ \ \ \ \ \ \ \ \ \ \ \ \ \ \ \ \ \ \ \ \ \ \ \ \ \ \ \ \ \ \ \ \ \ \ \ \ \ \ \ \ \ \ \ \ \ \ \ \ \ \ \ \ \ \ \ \ \ \ \ \ \ \ \ \ \ \ \ \ \ \ \ \ \ \ \ \ \ \ \ \ \ \ \ \ \ \ \ \ \ \ \ \ \ \ \ \ \ \ \ \ \ \ \ \ \ \ \ \ \ \ \ \ \ \ \ \ \ \ \ \ \ \ \ \ \ \ \ \ \ \ \ \ \ \ \ \ \ \ \ \ \ \ \ \ \ \ \ \ \ \ \ \ \ \ \ \ \ \ \ \ \ \ \ \ \ \ \ \ \ \ \ \ \ \ \ \ \ \ \ \ \ \ \ \ \ \ \ \ \ \ \ \ \ \ \ \ \ \ \ \ \ \ \ \ \ \ \ \ \ \ \ \ \ \ \ \ \ \ \ \ \ \ \ \ \ \ \ \ \ \ \ \ \ \ \ \ \ \ \ \ \ \ \ \ \ \ \ \ \ \ \ \ \ \ \ \ \ \ \ \ \ \ \ \ \ \ \ \ \ \ \ \ \ \ \ \ \ \ \ \ \ \ \ \ \ \ \ \ \ \ \ \ \ \ \ \ \ \ \ \ \ \ \ \ \ \ \ \ \ \ \ \ \ \ \ \ \ \ \ \ \ \ \ \ \ \ \ \ \ \ \ \ \ \ \ \ \ \ \ \ \ \ \ \ \ \ \ \ \ \ \ \ \ \ \ \ \ \ \ \ \ \ \ \ \ \ \ \ \ \ \ \ \ \ \ \ \ \ \ \ \ \ \ \ \ \ \ \ \ \ \ \ \ \ \ \ \ \ \ \ \ \ \ \ \ \ \ \ \ \ \ \ \ \ \ \ \ \ \ \ \ \ \ \ \ \ \ \ \ \ \ \ \ \ \ \ \ \ \ \ \ \ \ \ \ \ \ \ \ \ \ \ \ \ \ \ \ \ \ \ \ \ \ \ \ \ \ \ \ \ \ \ \ \ \ \ \ \ \ \ \ \ \ \ \ \ \ \ \ \ \ \ \ \ \ \ \ \ \ \ \ \ \ \ \ \ \ \ \ \ \ \ \ \ \ \ \ \ \ \ \ \ \ \ \ \ \ \ \ \ \ \ \ \ \ \ \ \ \ \ \ \ \ \ \ \ \ \ \ \ \ \ \ \ \ \ \ \ \ \ \ \ \ \ \ \ \ \ \ \ \ \ \ \ \ \ \ \ \ \ \ \ \ \ \ \ \ \ \ \ \ \ \ \ \ \ \ \ \ \ \ \ \ \ \ \ \ \ \ \ \ \ \ \ \ \ \ \ \ \ \ \ \ \ \ \ \ \ \ \ \ \ \ \ \ \ \ \ \ \ \ \ \ \ \ \ \ \ \ \ \ \ \ \ \ \ \ \ \ \ \ \ \ \ \ \ \ \ \ \ \ \ \ \ \ \ \ \ \ \ \ \ \ \ \ \ \ \ \ \ \ \ \ \ \ \ \ \ \ \ \ \ \ \ \ \ \ \ \ \ \ \ \ \ \ \ \ \ \ \ \ \ \ \ \ \ \ \ \ \ \ \ \ \ \ \ \ \ \ \ \ \ \ \ \ \ \ \ \ \ \ \ \ \ \ \ \ \ \ \ \ \ \ \ \ \ \ \ \ \ \ \ \ \ \ \ \ \ \ \ \ \ \ \ \ \ \ \ \ \ \ \ \ \ \ \ \ \ \ \ \ \ \ \ \ \ \ has
eigenstates $\widehat{u}_{\tau}f$ in the Schr\"{o}dinger-Weil representation
\[
\widehat{u}_{\tau}f=%
{\displaystyle\sum\limits_{k\in S;N(k)=\frac{\tau}{2\omega_{0}}}}
f(k)t_{x}^{k}I
\]
We observe that $\widehat{u}_{\sigma}$ projects out a function that is defined
on the spherical shell $N(k)=\frac{\tau}{2\omega_{0}}.$

We postpone a derivation of the pseudo-differential expressions for the
multiplicative operators $t_{s}^{a}$ and $t_{r}^{a+\sqrt{\delta}b}$ since this
entails some additional development. \ \ 

\section{Discussion}

\bigskip We have described the construction of Heisenberg groups H$_{\ast}(A)$
over an involutive ring $A$ and shown how this extends the usual construction
\cite{BS,Lo2}. \ One subgroup of automorphisms of this Heisenberg group
recovers the group SL$_{\ast}(2,A)$ described by J. Pantoja and J.
Soto-Andrade \cite{Pn}. \ Another subgroup of automorphisms of H$_{\ast}(A)$
are generated by elements from a diagonal matrix group D$(A_{0}^{x})$ whose
entries are from the set of norm 1 elements of $A.$ \ These automorphisms
preserve the norm of $A.$

We specialize to the case $\left(  \text{D}(A_{0}^{x})\otimes\text{SL}%
(2,F_{N})\right)  \ltimes$H$_{\ast}(S)$ for $S$ a subspace of the complex
quaternion group algebra $C4\otimes Q(F_{N})$ \ over the prime field $F_{N}$
$.$ \ Vector subspaces or subalgebras of $C4\otimes Q(F_{N})$ provide finite
models for each of the real or complex quadratic spaces with dimension 4 or
less. \ We construct a model that provides the Schr\"{o}dinger-Weil
representation of SL$(2,F_{N})\ltimes$H$_{\ast}(S)$ and indicate how it can be
extended to the complete group $\left(  \text{D}(A_{0}^{x})\otimes
\text{SL}(2,F_{N})\right)  \ltimes$H$_{\ast}(S).$ \ Functions that transform
according to the Schr\"{o}dinger representation of H$_{\ast}(S)$ can be
associated with wave-functions of quantum systems having $S$ as configuration
space. \ The evolution of the wavefunction under quadratic Hamiltonians is
described by the Weil representation of its automorphism group \cite{deG}. \ 

We define a pseudo-differential operator that can be applied to functions in
the group algebra of a finite translation group. \ We obtain differential
expressions for particular group actions in the Schr\"{o}dinger-Weil
representation of $\left(  \text{D}(A_{0}^{x})\otimes\text{SL}(2,F_{N}%
)\right)  \ltimes$H$_{\ast}(S)$. \ \ The pseudo-differential operator enables
a parallel treatment of spaces defined over finite and real fields.

\appendix

\section{Action of j:}

\bigskip We determine the action of $j$ by direct calculation. \ The full
calculation, though elementary, is long. \ We do the first portion below and
then indicate key intermediate results towards the final result. \ The
complete calculation for the 1 dimensional case is contained in the appendix
of Johnson \cite{J2}.

We use the identity%
\begin{equation}
t_{x}^{k}\widehat{y}_{\sigma}\widehat{z}_{\omega_{0}}=\widehat{y}%
_{\sigma-2\omega_{0}k^{\ast}}\widehat{z}_{\omega_{0}}t_{x}^{k} \label{ident}%
\end{equation}
that is derived by expanding out $\widehat{y}_{\sigma}$ according to the
definition Eq. \ref{estate for tx} and passing $t_{x}^{k}$ through the
expression using the commutator Eq. \ref{h commutator}.

\bigskip Consider the action of $j$ on $f\ $\ given by Eq. \ref{swrep1}%

\[
jf=j%
{\displaystyle\sum\limits_{k\in S}}
f(k)t_{x}^{k}\widehat{y}_{0}\widehat{z}_{\omega_{0}}\widehat{s}_{\chi}%
\widehat{u}_{0}\_,
\]
where, to simplify notation, the underline in the preceding expression denotes
all of the terms following $\widehat{u}_{0}.$ \ Expand out $\widehat{y}_{0}$
using $\widehat{y}_{0}=\frac{1}{V}%
{\displaystyle\sum\limits_{h\in S}}
t_{y}^{h}$ and act on the resulting term $t_{x}^{k}t_{y}^{h}$ with $j$ using
$j\left(  t_{x}^{p}t_{y}^{q}t_{z}^{\text{tr}(t)}\right)  j^{-1}=t_{x}^{q}%
t_{y}^{-p}t_{z}^{\text{tr}(t+pq^{\ast})}.$ Noting that $j\widehat{s}_{\chi
}=\widehat{s}_{\chi}j$ for the cases $\chi=0,\frac{N-1}{2}$ that are of
interest here, we obtain%
\[
jf=%
{\displaystyle\sum\limits_{k\in S}}
f(k)\frac{1}{V}%
{\displaystyle\sum\limits_{h\in S}}
t_{x}^{h}t_{y}^{-k}t_{z}^{\text{tr}(kh^{\ast})}\widehat{z}_{\omega_{0}%
}\widehat{s}_{\chi}j\widehat{u}_{0}\_
\]
Now insert $1=%
{\displaystyle\sum\limits_{\sigma\in S}}
\widehat{y}_{\sigma}$%
\[
jf=%
{\displaystyle\sum\limits_{k\in S}}
f(k)\frac{1}{V}%
{\displaystyle\sum\limits_{h\in S}}
t_{x}^{h}t_{y}^{-k}t_{z}^{\text{tr}(kh^{\ast})}%
{\displaystyle\sum\limits_{\sigma\in S}}
\widehat{y}_{\sigma}\widehat{z}_{\omega_{0}}\widehat{s}_{\chi}j\widehat{u}%
_{0}\_
\]
Let $t_{y}^{-k}$ act on\ $\widehat{y}_{\sigma}$ and $t_{z}^{\text{tr}%
(kh^{\ast})}$ act on $\widehat{z}_{\omega_{0}}$ to obtain the exponential term%
\[
\exp(\frac{-2\pi i}{N}\text{Tr}k(\sigma-2\omega_{0}h^{\ast})).
\]
Insert $1=t_{x}^{\frac{-\sigma^{\ast}}{2\omega_{0}}}t_{x}^{\frac{\sigma^{\ast
}}{2\omega_{0}}}$ to the left of $\widehat{y}_{\sigma}.$ \ We obtain%
\[
jf=%
{\displaystyle\sum\limits_{k\in S}}
f(k)\frac{1}{V}%
{\displaystyle\sum\limits_{h,\sigma\in S}}
\exp(\frac{-2\pi i}{N}\text{Tr}[k(\sigma-2\omega_{0}h^{\ast})])t_{x}^{h}%
t_{x}^{\frac{-\sigma^{\ast}}{2\omega_{0}}}t_{x}^{\frac{\sigma^{\ast}}%
{2\omega_{0}}}\widehat{y}_{\sigma}\widehat{z}_{\omega_{0}}\widehat{s}_{\chi
}j\widehat{u}_{0}\_
\]
Now factor $t_{x}^{\frac{\sigma^{\ast}}{2\omega}}$ through $\widehat
{y}_{\sigma}$ and use the identity Eq. \ref{ident} to obtain%
\[
jf=%
{\displaystyle\sum\limits_{k\in S}}
f(k)\frac{1}{V}%
{\displaystyle\sum\limits_{h,\sigma\in S}}
\exp(\frac{-2\pi i}{N}\text{Tr}k(\sigma-2\omega_{0}h^{\ast}))t_{x}%
^{\frac{h2\omega_{0}-\sigma^{\ast}}{2\omega_{0}}}\widehat{y}_{0}\widehat
{z}_{\omega_{0}}t_{x}^{\frac{\sigma^{\ast}}{2\omega_{0}}}\widehat{s}_{\chi
}j\widehat{u}_{0}\_
\]
Substitute $h^{\prime}=\frac{h2\omega_{0}-\sigma^{\ast}}{2\omega_{0}}$ and sum
over $\sigma.$ \ We obtain%
\begin{align}
jf  &  =%
{\displaystyle\sum\limits_{h^{\prime}\in S}}
\left(
{\displaystyle\sum\limits_{k\in S}}
f(k)\exp(\frac{2\pi i}{N}2\omega_{0}\text{Tr}kh^{\prime\ast})\right)
t_{x}^{h^{\prime}}\cdot\nonumber\\
&  \cdot\widehat{y}_{0}\widehat{z}_{\omega_{0}}\widehat{s}_{\chi}\left(
\widehat{x}_{0}j\right)  \widehat{u}_{0}\left\{
{\displaystyle\sum\limits_{p\in S}}
t_{x}^{p}\left(  1+\tau%
{\displaystyle\sum\limits_{l\in S;N(l)=0}}
\exp(\frac{2\pi i}{N}2\omega_{0}\text{Tr}lp^{\ast})\right)  \right\}
(1+\alpha j)\widehat{d}_{0}\nonumber\\
&  \label{jintermed}%
\end{align}

It remains to show that the second line of Eq. \ref{jintermed} is equal, up to
a prefactor, to Eq. \ref{ideal} for chosen values of $\tau,\alpha$ and
$\widehat{s}_{\chi}.$ \ In outline, we pass $\left(  \widehat{x}_{0}j\right)
$ to the right through the second line of expression Eq. \ref{jintermed} and
\ compare with Eq. \ref{swrep1}. \ We find three types of terms: \ terms that
equate to the $\alpha j$ portion of Eq. \ref{swrep1} up to a prefactor, terms
that sum to zero, and terms that equate to the non-$\alpha j$ portion of Eq.
\ref{swrep1} up to a prefactor. \ The first two types provide conditions on
$\chi,\tau$ and $\alpha.$ \ We then find that the condition arising from the
non-$\alpha j$ portion is satisfied automatically.

The parameters $\tau$ and $\alpha$ can be expressed in terms of 2 sums%
\begin{align*}
c_{0}  &  =%
{\displaystyle\sum\limits_{l\in S;N(l)=0}}
1\\
c_{1}  &  =%
{\displaystyle\sum\limits_{l\in S;N(l)=0}}
\exp(\frac{2\pi i}{N}2\omega_{0}\text{Tr}lh^{\ast})
\end{align*}
for fixed $h$ such that $h\neq0,N(h)=0$. $\ \tau$ satisfies the quadratic
equation%
\[
0=\tau^{2}(c_{0}-c_{1})-\tau c_{1}-1
\]
and $\alpha$ can be expressed%
\[
\alpha=\frac{1}{N\tau(c_{0}-c_{1})}G(1,N)^{\dim}\left(  \frac{-2\omega_{0}}%
{N}\right)  ^{\dim}\left(  \frac{-1}{N}\right)  ^{q}%
\]
where $\dim$ is the dimensionality of $S,$ $G(1,N)\ $is the Gauss sum, and $q$
is the number of negative terms in the quadratic form. \ The representation
$\widehat{s}_{\chi}$ is the trivial representation $\widehat{s}_{+}$ when
$\dim$ is even and the representation $\widehat{s}_{-}$ when $\dim$ is odd. \ \ 

For all cases we can express the action of $j$ as%

\begin{equation}
jf=\kappa%
{\displaystyle\sum\limits_{h\in S}}
\left(
{\displaystyle\sum\limits_{k\in S}}
f(k)\exp(\frac{2\pi i}{N}2\omega_{0}\text{Tr}kh^{\ast})\right)  t_{x}^{h}I
\label{jact}%
\end{equation}
for%
\[
\kappa=\frac{1}{V}G(1,N)^{\dim}\left(  \frac{2\omega_{0}}{N}\right)  ^{\dim
}\left(  \frac{-1}{N}\right)  ^{q}%
\]
\ 

The trace of the linear transformation Eq. \ref{jact} is the character of $j$.
\ It has value $\left(  \frac{-2}{N}\right)  ^{\dim}$ (cf., \cite{N,Th}).

\end{document}